\documentclass[conference]{IEEEtran}

\usepackage{graphicx}
\usepackage{epstopdf}
\usepackage{standalone}
\usepackage[nolist ]{acronym}
\usepackage{tcolorbox}
\usepackage{colortbl}

\usepackage{pdfpages}
\usepackage{hyperref}
\usepackage{pdfcomment}
\usepackage{hologo}

\usepackage{xstring}

\usepackage[utf8]{inputenc}
\usepackage{marvosym}

\usepackage{algorithm}
\usepackage{algpseudocode}
\usepackage{tikz}
\usetikzlibrary{positioning}

\usepackage{listings}
\definecolor{ListingBackground}{rgb}{0.97,0.97,0.97}
\usepackage{pgfplots}
\pgfplotsset{compat=newest}
\pgfplotsset{
    box plot/.style={
        /pgfplots/.cd,
        fill=blue!30,
        only marks,
        mark=-,
        mark size=0.2em,
        /pgfplots/error bars/.cd,
        y dir=plus,
        y explicit,
    },
    box plot box/.style={
        /pgfplots/error bars/draw error bar/.code 2 args={%
            \draw  ##1 -- ++(.2em,0pt) |- ##2 -- ++(-.2em,0pt) |- ##1 -- cycle;
        },
        /pgfplots/table/.cd,
        y index=2,
        y error expr={\thisrowno{3}-\thisrowno{2}},
        /pgfplots/box plot
    },
    box plot top whisker/.style={
        /pgfplots/error bars/draw error bar/.code 2 args={%
            \pgfkeysgetvalue{/pgfplots/error bars/error mark}%
            {\pgfplotserrorbarsmark}%
            \pgfkeysgetvalue{/pgfplots/error bars/error mark options}%
            {\pgfplotserrorbarsmarkopts}%
            \path ##1 -- ##2;
        },
        /pgfplots/table/.cd,
        y index=4,
        y error expr={\thisrowno{2}-\thisrowno{4}},
        /pgfplots/box plot
    },
    box plot bottom whisker/.style={
        /pgfplots/error bars/draw error bar/.code 2 args={%
            \pgfkeysgetvalue{/pgfplots/error bars/error mark}%
            {\pgfplotserrorbarsmark}%
            \pgfkeysgetvalue{/pgfplots/error bars/error mark options}%
            {\pgfplotserrorbarsmarkopts}%
            \path ##1 -- ##2;
        },
        /pgfplots/table/.cd,
        y index=5,
        y error expr={\thisrowno{3}-\thisrowno{5}},
        /pgfplots/box plot
    },
    box plot median/.style={
        /pgfplots/box plot
    },
    boxplot/every median/.style={
    	ultra thick,dashed,cyan
    }
}

\definecolor{flexicolor}{RGB}{46,49,146}
\definecolor{amaricolor}{RGB}{237,28,36}

\usepackage{xspace}

\usepackage[binary-units=true]{siunitx}
\usepackage{psfrag}
\usepackage{graphicx}
\usepackage{tabularx,booktabs}
\usepackage{multirow}
\usepackage{rotating}

\renewcommand{\baselinestretch}{1}

\usepackage{multirow}

\usepackage[cmex10]{amsmath}
\usepackage[caption=false,font=footnotesize]{subfig}
\usepackage{stfloats}
\renewcommand{\baselinestretch}{0.93}

\begin{document}

\newcommand{\paperTitle}{Empirical Analysis of Client-based Network Quality Prediction in Vehicular Multi-MNO Networks}
\newcommand{\paperAuthors}{Benjamin Sliwa and Christian Wietfeld}
\newcommand{\paperEmails}{$\{$Benjamin.Sliwa, Christian.Wietfeld$\}$@tu-dortmund.de}

\newcommand{\figurePadding}{0pt}
\newcommand{\figureTopPadding}{\figurePadding}
\newcommand{\figureBottomPadding}{\figurePadding}

\newcommand{\double}{0.45\textwidth}
\newcommand{\triple}{0.32\textwidth}
\newcommand{\quarter}{0.23\textwidth}

\newcommand{\mnoA}{\emph{\mno~A}\xspace}
\newcommand{\mnoB}{\emph{\mno~B}\xspace}
\newcommand{\mnoC}{\emph{\mno~C}\xspace}

\newcommand{\ctext}[3][RGB]{%
	\begingroup
	\definecolor{hlcolor}{#1}{#2}\sethlcolor{hlcolor}%
	\hl{#3}%
	\endgroup
}

\newcommand\keyResult[1]{
	\ifx\KEYRESULTS\VOID
		#1
	\else
		\colorbox{yellow}{\parbox{1\columnwidth}{#1}}
	\fi
}

\newcommand{\dummy}[3]
{
	\begin{figure}[b!]  
		\begin{tikzpicture}
		\node[draw,minimum height=6cm,minimum width=\columnwidth]{\LARGE #1};
		\end{tikzpicture}
		\caption{#2}
		\label{#3}
	\end{figure}
}

\newcommand{\wDummy}[3]
{
	\begin{figure*}[b!]  
		\begin{tikzpicture}
		\node[draw,minimum height=6cm,minimum width=\textwidth]{\LARGE #1};
		\end{tikzpicture}
		\caption{#2}
		\label{#3}
	\end{figure*}
}

\newcommand{\basicFig}[7]
{
	\begin{figure}[#1]  	
		\vspace{#6}
		\centering		  
		\includegraphics[width=#7\columnwidth]{#2}
		\caption{#3}
		\label{#4}
		\vspace{#5}	
	\end{figure}
}
\newcommand{\fig}[4]{\basicFig{#1}{#2}{#3}{#4}{0cm}{0cm}{1}}

\newcommand{\subfig}[3]
{
	\subfloat[#3]{\includegraphics[width=#2\textwidth]{#1}}\hfill%
}

\newcommand\circled[1] % caution with using in captions: \protect \circled
{
	\tikz[baseline=(char.base)]
	{
		\node[shape=circle,draw,inner sep=1pt] (char) {#1};
	}\xspace
}

\newcommand{\sideHeader}[3]
{
	\multirow{#1}{*}{
		\rotatebox[origin=c]{90}{
			\parbox{#2}{\centering \textbf{#3}}
		}
	}
}
\begin{acronym}
	\acro{CoPoMo}{Context-aware Power Consumption Model}
	\acro{KPI}{Key Performance Indicator}
	\acro{LTE}{Long Term Evolution}
	\acro{eNB}{evolved Node B}
	\acro{UE}{User Equipment}
	\acro{RTT}{Round Trip Time}
	\acro{TCP}{Transmission Control Protocol}
	\acro{UDP}{User Datagram Protocol}
	\acro{V2X}{Vehicle-to-everything}
	\acro{MNO}{Mobile Network Operator}
	
	\acro{ASU}{Arbitrary Strength Unit} 
	\acro{RSRP}{Reference Signal Received Power} 
	\acro{RSRQ}{Reference Signal Received Quality} 
	\acro{SINR}{Signal-to-interference-plus-noise Ratio} 
	\acro{CQI}{Channel Quality Indicator} 
	\acro{TA}{Timing Advance} 
	
	\acro{ANN}{Artificial Neural Network} 
	\acro{LR}{Linear Regression} 
	\acro{SVM}{Support Vector Machine} 
	\acro{M5}{M5 Regression Tree} 
	\acro{RF}{Random Forests} 
	\acro{MAE}{Mean Absolute Error} 
	\acro{WEKA}{Waikato Environment for Knowledge Analysis}
	\acro{URLLC}{Ultra-reliable Low Latency Communication}
	\acro{CM}{Connectivity Map}
	\acro{GPS}{Global Positioning System}
	\acro{MDI}{Mean Decrease Impurity}
	\acro{ECDF}{Empirical Cumulative Distribution Function}
	\acro{mmWave}{millimeter Wave}
	\acro{CART}{Classification And Regression Tree}
\end{acronym}

\newcommand\copomo{\ac{CoPoMo}\xspace}
\newcommand\kpi{\ac{KPI}\xspace}
\newcommand\kpis{\acp{KPI}\xspace}
\newcommand\lte{\ac{LTE}\xspace}
\newcommand\enb{\ac{eNB}\xspace}
\newcommand\ue{\ac{UE}\xspace}
\newcommand\tcp{\ac{TCP}\xspace}
\newcommand\udp{\ac{UDP}\xspace}
\newcommand\mno{\ac{MNO}\xspace}
\newcommand\mnos{\acp{MNO}\xspace}
\newcommand\rtt{\ac{RTT}\xspace}
\newcommand\rsrp{\ac{RSRP}\xspace}
\newcommand\rsrq{\ac{RSRQ}\xspace}
\newcommand\sinr{\ac{SINR}\xspace}
\newcommand\cqi{\ac{CQI}\xspace}
\newcommand\ann{\ac{ANN}\xspace}
\newcommand\lr{\ac{LR}\xspace}
\newcommand\svm{\ac{SVM}\xspace}
\newcommand\rf{\ac{RF}\xspace}
\newcommand\mae{\ac{MAE}\xspace}
\newcommand\gps{\ac{GPS}\xspace}
\newcommand\mmWave{\ac{mmWave}\xspace}
\acresetall
\title{\paperTitle}

\author{\IEEEauthorblockN{\textbf{\paperAuthors}}
	\IEEEauthorblockA{Communication Networks Institute, TU Dortmund University, 44227 Dortmund, Germany\\
		e-mail: \paperEmails}}

\maketitle

%
% Make your adjustments here
%
\def\COPYRIGHTYEAR{2019}
\def\CONFERENCE{2019 IEEE 90th IEEE Vehicular Technology Conference (VTC-Fall)} % set after the paper has been accepted
\def\DOI{10.1109/VTCFall.2019.8891392}	% set after the paper has been published

\def\bibtex
{
	@InProceedings\{Sliwa/Wietfeld/2019b,\\
	author    = \{Benjamin Sliwa and Christian Wietfeld\},\\
	title     = \{Empirical analysis of client-based network quality prediction in vehicular multi-\{MNO\} networks\},\\
	booktitle = \{2019 IEEE 90th Vehicular Technology Conference (VTC-Fall)\},\\
	year      = \{2019\},\\
	address   = \{Honolulu, Hawaii, USA\},\\
	month     = \{Sep\},\\
	\}
}
\ifx\CONFERENCE\VOID
\def\conferencenotice{Submitted for publication}
\def\copyrightnotice{}
\else
\ifx\DOI\VOID
\def\conferencenotice{Accepted for presentation in: \CONFERENCE}	
\else
\def\conferencenotice{Published in: \CONFERENCE\\DOI: \href{http://dx.doi.org/\DOI}{\DOI}

}
\fi
\def\copyrightnotice{
	\copyright~\COPYRIGHTYEAR~IEEE. Personal use of this material is permitted. Permission from IEEE must be obtained for all other uses, including reprinting/republishing this material for advertising or promotional purposes, collecting new collected works for resale or redistribution to servers or lists, or reuse of any copyrighted component of this work in other works.
}
\fi
\def\overlayimage{%
	\begin{tikzpicture}[remember picture, overlay]
		\node[below=5mm of current page.north, text width=20cm,font=\sffamily\footnotesize,align=center] {\conferencenotice \vspace{0.3cm} \pdfcomment[color=yellow,icon=Note]{\bibtex}};
	\node[above=5mm of current page.south, text width=15cm,font=\sffamily\footnotesize] {\copyrightnotice};
	\end{tikzpicture}%
}
\overlayimage
\begin{abstract}
	
%
% Introduction
%
Multi-\mno networking is a promising method to exploit the joint force of multiple available cellular data connections within vehicular networks. By applying anticipatory communication principles, data transmissions can dynamically utilize the mobile network with the highest estimated network performance in order to achieve improvements in data rate, resource efficiency, and reliability.

%
% Solution appraoch
%
In this paper, we present the results of a comprehensive real-world measurement campaign in public cellular networks in different scenarios and analyze the performance of online data rate prediction based on multiple machine learning models and data aggregation strategies.	
%
% Benefits
%
It is shown that multi-\mno approaches are able to achieve significant benefits for all considered network quality and end-to-end indicators even in the presence of a single dominant \mno.
%
% Prediction
%
However, the analysis points out that anticipatory multi-\mno communication requires the consideration of \mno-specific machine learning models since the impact of the different features is highly depending on the configuration of the network infrastructure.

\end{abstract}

\IEEEpeerreviewmaketitle

\section{Introduction}

%
% Introduction
%
Anticipatory communication \cite{Bui/etal/2017a} has been proposed as a method to face the various challenges of next-generation networks, which are related to higher requirements (e.g., latency and data rate as enablers for autonomous driving) as well as massive increases in the overall data usage \cite{Sliwa/etal/2019b}.
%
% Properties and Examples
%
This novel communication paradigm aims to optimize decision processes within communication networks by proactively integrating context information such as mobility predictions and network quality parameters. Applications range from opportunistic sensor data transfer for optimizing the coexistence of different cell users \cite{Sliwa/etal/2018b} to reliable and mobility-aware multi-hop routing \cite{Sliwa/etal/2016b}.
%
% Requirements for Anticipatory Communication
%
Since anticipatory communications is a \emph{data-driven} approach, it relies on knowledge and prediction models. In recent work \cite{Sliwa/etal/2018a, Sliwa/etal/2019d}, we have demonstrated the potentials of using machine learning-based data rate prediction and multi-layer connectivity maps for improving the resource-efficiency of vehicular sensor data transmissions. Furthermore, we have shown that the prediction models can be exploited for analyzing novel communication methods in close to reality scenarios \cite{Sliwa/Wietfeld/2019c}. 
In this paper, we provide an empirical analysis of the potentials of using data rate prediction within multi-connectivity cellular vehicular networks. The real world evaluation is carried out in the public cellular networks of the three \mnos in Germany and paves the way for future development of context-aware multi-connectivity optimizations. It furthermore shows how upcoming 5G-based vehicular communication systems can exploit multi-connectivity for compensating coverage gaps.

Fig.~\ref{fig:rsrp_time} shows an example trace of the \lte \rsrp for three different \mnos during a drive test in a suburban environment. It can be seen that although not all carriers are able to guarantee seamless connectivity for the whole experiment duration, a multi-carrier approach is able to do so.
%
% Fig. RSRP vs Time
%
\basicFig{b}{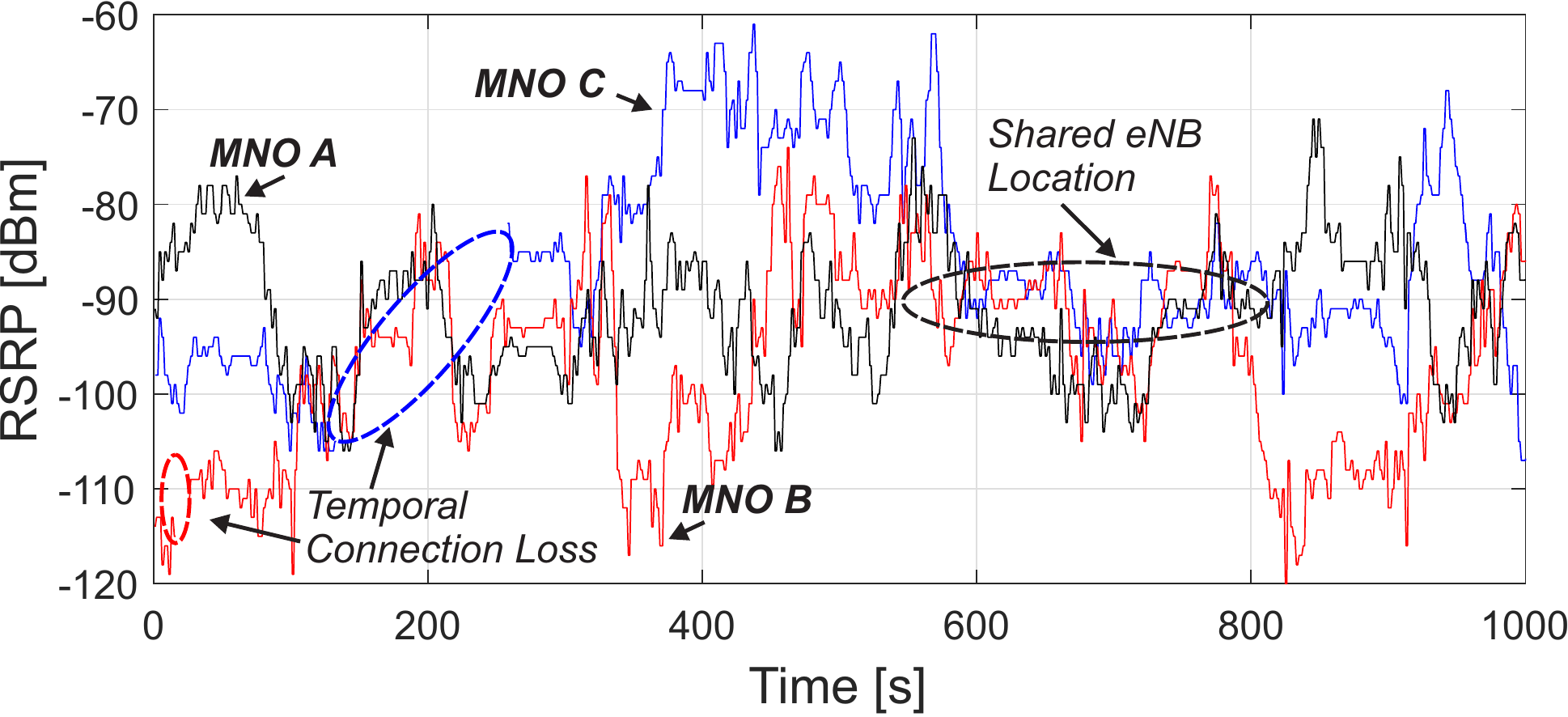}{Example \rsrp trace for three different network carriers in suburban environment. Although individual networks fail to provide connectivity at certain locations, the combined approach is able to provide full coverage.}{fig:rsrp_time}{0cm}{-0.4cm}{1}
%
% Structure of the paper
%
The contributions provided by this paper are as follows:
\begin{itemize}
	\item Comprehensive \textbf{real world performance evaluation of vehicular multi-\mno communication} in the public cellular \lte networks of German \mnos.
	\item Statistical analysis of the potential benefits by combining networks of multiple \mnos and data management techniques using multi-\mno \acp{CM}
	\item Performance evaluation of \textbf{client-based online data rate prediction} with different data aggregation methods and machine learning models.	
	\item For achieving a high grade of transparency and reproducibility, all raw results of the measurements and the measurement application itself are provided in an \textbf{Open Access} way.
\end{itemize}

%
% Structure
%
The remainder of the paper is structured as follows. After discussing relevant approaches for machine learning-based data rate prediction and evaluations of multi-connectivity approaches in Sec.~\ref{sec:related_work}, we present the methodological model for the empirical performance evaluation in Sec.~\ref{sec:methods}. 
%
% Data
%
The result analysis is divided into two parts. At first, the statistical properties of the network quality indicators, potential benefits of using multi-\mno connectivity as well as methods for maintaining multi-\mno data are discussed in Sec.~\ref{sec:data}.
%
% Machine Learning
%
Afterwards, machine learning-based data rate prediction is discussed in Sec.~\ref{sec:prediction}. The analysis includes the application of different prediction models in uplink and downlink direction as well as data aggregation and feature importance evaluations.

\section{Related Work} \label{sec:related_work}

Since anticipatory communication highly relies on predicting unobserved values from measurable information, it is closely related to machine learning \cite{Ye/etal/2018a}. A comprehensive overview about different machine learning models and their application fields in wireless communication networks is provided by \cite{Jiang/etal/2017a}.

%
% Network Quality
%
Within vehicular networks, the performance of end-to-end indicators such as throughput and latency is highly impacted by the encountered network quality \cite{Walelgne/etal/2018a, Akselrod/etal/2017a}. Therefore, many research works investigate passive data rate prediction based on measured network quality indicators, e.g. for interface selection \cite{Cavalcanti/etal/2018a}. A survey about different bandwidth estimation techniques is provided by \cite{Chaudhari/Biradar/2015a}.

The authors of \cite{Riihijarvi/Mahonen/2018a} evaluate different machine learning models (\acp{ANN}, \ac{LR} and \rf) for predicting the downlink data rate on a large data set using signal quality measurements only. They conclude that classical machine learning methods yield excellent prediction results -- with the \rf model achieving the lowest error probability -- for providing the \mno with additional information. However since no information about the payload size about the transmitted packets is provided, it can be assumed that a fixed size was used for the experiments. As other work has shown \cite{Sliwa/etal/2018b, Sliwa/etal/2018a, Sliwa/etal/2019d} the payload size is an important feature that needs to be considered in order to derive models, which generalize well on other data sets.
A simple linear regression approach based on OpenSignal data is proposed in \cite{Cainey/etal/2014a}.
While the authors point out that the \rsrp seems to have the dominant effect on prediction accuracy, they argue that a better compensation of multipath-effects could be achieved by integrating additional indicators (e.g., \rsrq, \sinr) into the prediction.
%
% Fig. KPI Histograms Campus
%
\begin{figure*}[b] 
	\centering
	%
	% Campus
	%
	\includegraphics[width=\quarter]{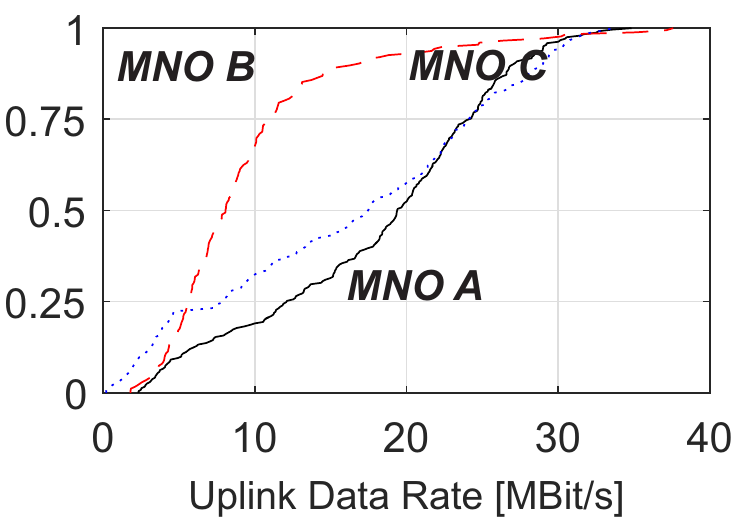}\quad
	\includegraphics[width=\quarter]{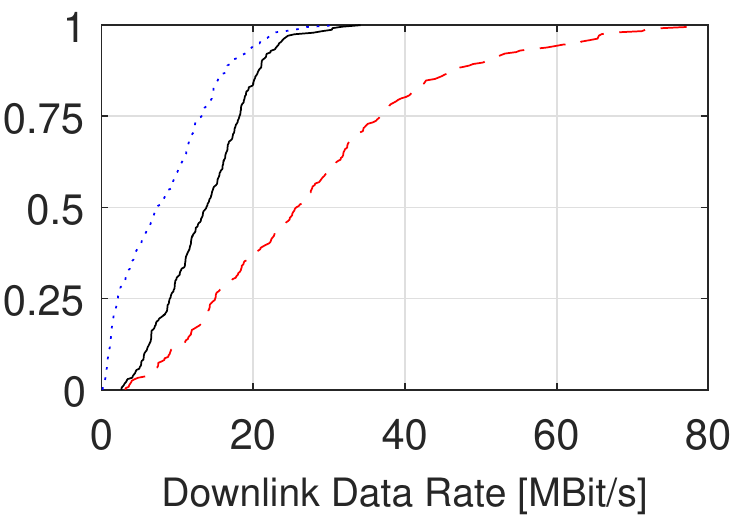}\quad
	\includegraphics[width=\quarter]{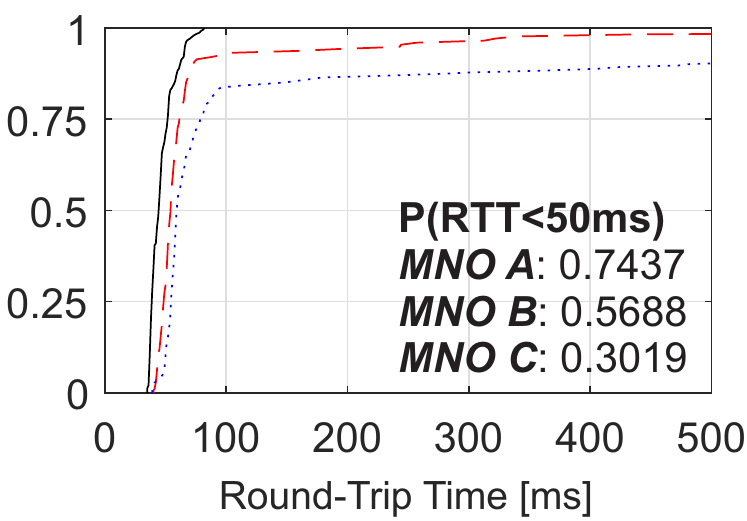}\quad
	\includegraphics[width=\quarter]{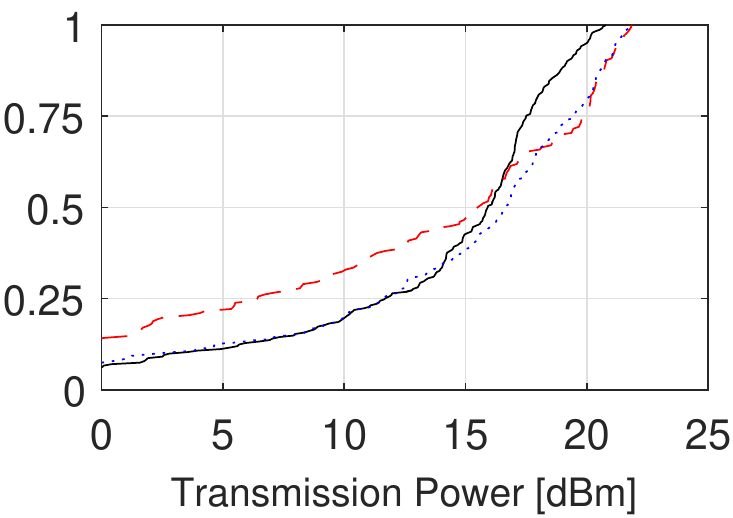}
	\vspace{-0.2cm}
	\caption{\emph{Urban} scenario: \acp{ECDF} of different indicators for the considered \acp{MNO}.}
	\label{fig:kpi_histograms_campus}
	\vspace{-0.2cm}
\end{figure*}
%
% Fig. KPI Histograms Highway
%
\begin{figure*}[b] 
	\centering
	%
	% Highway
	%
	\includegraphics[width=\quarter]{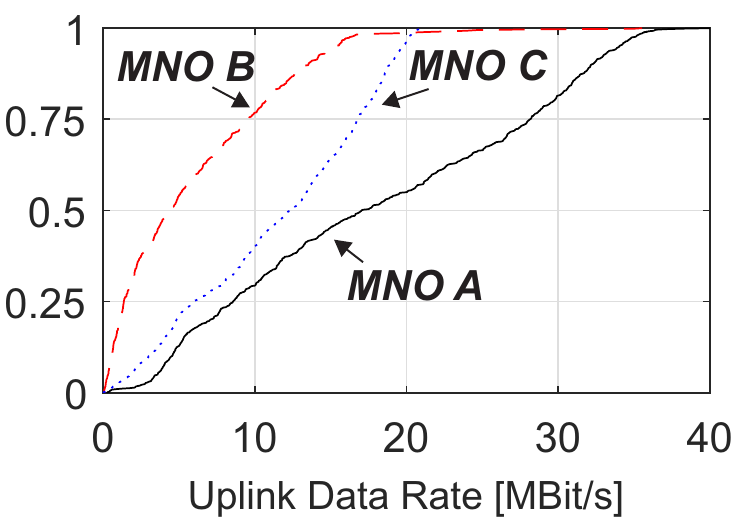}\quad
	\includegraphics[width=\quarter]{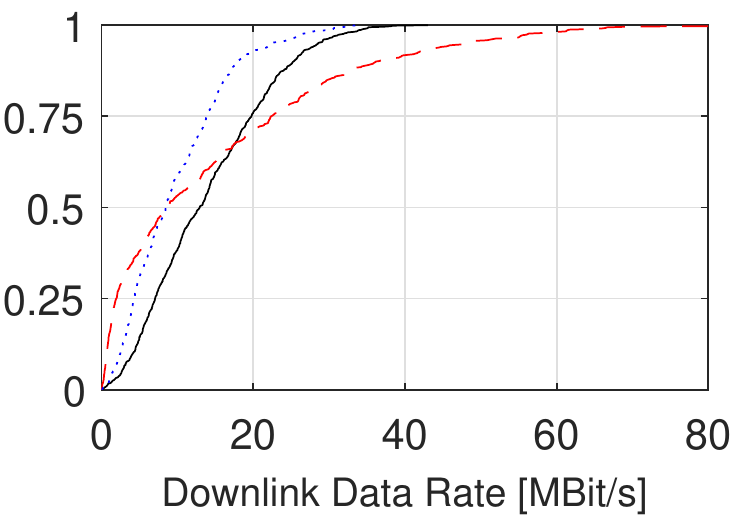}\quad
	\includegraphics[width=\quarter]{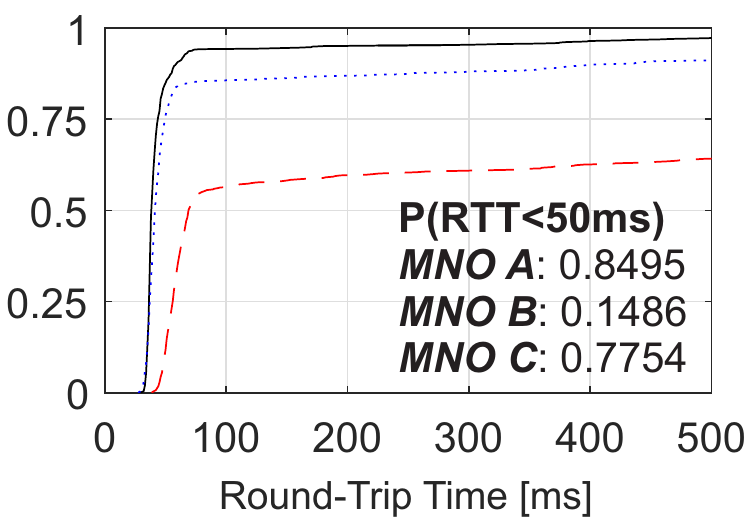}\quad
	\includegraphics[width=\quarter]{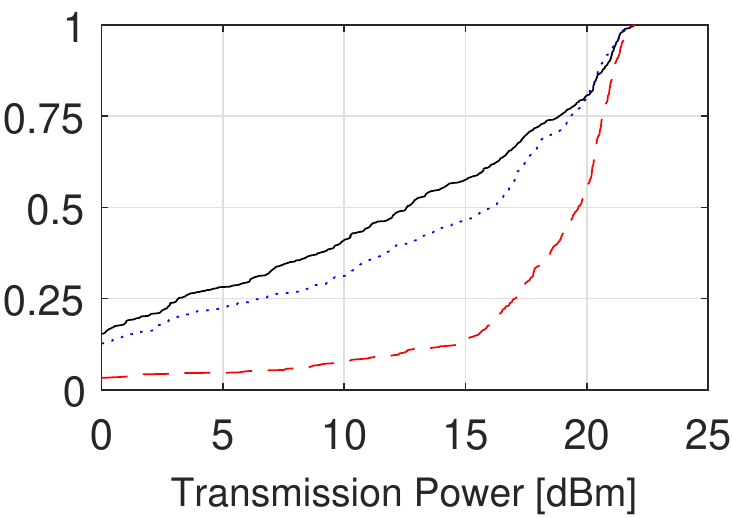}
	\vspace{-0.2cm}
	\caption{\emph{Highway} scenario: \acp{ECDF} of different indicators for the considered \acp{MNO}.}
	\label{fig:kpi_histograms_highway}
\end{figure*}
%
% Jomrich
%
The authors of \cite{Jomrich/etal/2018a} present a real world measurement campaign for cellular bandwidth estimation in vehicular environments for two different \mnos. While the achieved prediction accuracy is relatively low due to the focus on solely radio-related features, it is shown that the observed behavior differs depending on the \mno. Similar to the work discussed in \cite{Samba/etal/2017a}, the \rf model achieves the best prediction accuracy.
%
% Multi-connectivity
%
This fact is confirmed by \cite{Kousias/etal/2017a} Kousias et al., who point out that the relevance of the individual features is highly depending on the \mno. Furthermore, daytime-specific aspects are evaluated.
%
% UE-based Estimation of Available Uplink Data Rates
%
In \cite{Nikolov/etal/2018a}, the authors analyze passive \lte data rate prediction as a metric for interface selection within highly mobile networks. While the considered \ann approach achieves the highest accuracy, the authors present a \lr-based approach as a lightweight alternative for online application. However, the size of the training set is low and the authors do not consider the packet size within their prediction mechanism.

%
% Connectivity Maps
%
Although vehicles are able to measure the current channel context and can apply methods to estimate their trajectories, they are not able to directly measure the network quality at future locations. Connectivity maps \cite{Kasparick/etal/2016a} provide a way to map location data to network quality information. They are often created using a crowdsensing approach and are utilized as a priori information \cite{Apajalahti/etal/2018a}, allowing to make network quality forecasts based on previous measurement in the same geospatial area \cite{Sliwa/etal/2018a}.

\section{Methodology of the Empirical Evaluation} \label{sec:methods}

In this section, the methodological aspects of the measurement campaign are explained.
For the empirical performance evaluation, drive tests are performed in four scenarios with different velocity profiles and building densities: \emph{campus} (3~km), \emph{urban} (3~km), \emph{suburban} (9~km), \emph{highway} (14~km), whereas each track is evaluated ten times.

During the drive tests, three Android-based \acp{UE} (Samsung Galaxy S5 Neo, Model SM-G903F), which use the cellular link of one of the considered \mnos, are used to capture the network quality indicators and perform the transmissions for data rate and \rtt evaluations. 
%
% Data Transfer
%
Each 10~s, the measurement application\footnote{Measurement software available at https://github.com/BenSliwa/MTCApp} performs \tcp transmissions in uplink and downlink direction with variable payload sizes in the range of 0.1~MB to 10~MB. The data transfer is performed in the public cellular network and the throughput is calculated on the server side. \rtt is measured on the transport layer. Transmissions are only performed if a valid \lte connection is present.
%
% Statistics
%
In total, 12938 transmissions (58.45~GB) are performed on a total driven distance of 287~km. The raw real world measurements, which consist of time series data for trajectories and network quality indicators, can be accessed via \cite{Sliwa/2019a}.

%
% Data Analysis
%
For performing the machine learning-related evaluation, \ac{WEKA} \cite{Hall/etal/2009a} and \texttt{LIBSVM} \cite{Chang/Lin/2011a} are used.

%
% Fig. Connectivity Map
%
\begin{figure*}[] 
	\centering
	\includegraphics[width=1\textwidth]{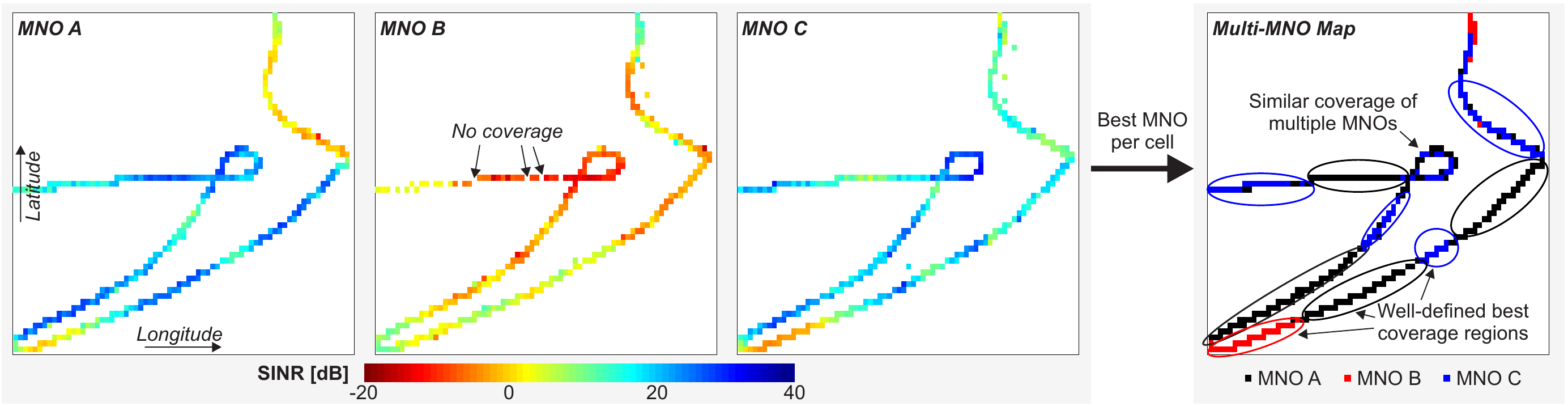}
	\vspace{-0.5cm}
	\caption{Excerpt of the connectivity maps of the highway track for all considered \mnos. The average \sinr per cell is shown as an example network quality indicator. The operator map visualizes the \mno with the highest \sinr per cell.}
	\label{fig:connectivityMap}
	\vspace{-0.5cm}
\end{figure*}

\section{Statistical Analysis of Multi-\mno Vehicular Networking} \label{sec:data}

In this section, the results and network quality characteristics are analyzed from a statistical perspective. In addition, potential benefits of using multi-\mno communication in vehicular networks are discussed.

\subsection{Statistical Properties} 
Due to spacial constraints, we focus on comparing two example scenarios \emph{campus} (see Fig.~\ref{fig:kpi_histograms_campus}) and \emph{highway} (see Fig.~\ref{fig:kpi_histograms_highway}). The figures show the \acp{ECDF} for data rates in uplink and downlink direction,  \rtt and applied transmission power $\text{P}_{\text{TX}}$.
%
% Throughput
%
While \mnoA provides a similar network performance in both scenarios, the results for \mnoB and \mnoC are highly scenario-dependent and directly related to the \lte coverage (cf. Tab.~\ref{tab:multiMNO}).
%
% RTT
%
In the highway scenario, the overall \lte coverage of \mnoC is only $92~\%$. Therefore, multiple handovers to 3G occured during active transmissions, which is the reason for the high \rtt and the low throughput of \mnoB. Note that \rtt is measured on the transport layer and is therefore impacted by the network quality and \tcp-related features such as retransmissions.
%
% Transmission Power
%
Due to the low coverage, the average distance to the serving \enb is high. Therefore, the \ue aims to compensate the path loss by increasing the transmission power $\text{P}_{\text{TX}}$, which is much higher for \mnoB than for the other considered \mnos. As discussed in \cite{Falkenberg/etal/2018a}, $\text{P}_{\text{TX}}$ has a dominant impact on the overall power consumption of the \ue. In the highway scenario, it can be expected that using \mnoB leads to a significantly shorter battery lifetime, which also fits our qualitative observations during the measurements.

\subsection{Multilayer Multi-\mno Connectivity Maps}

The acquired data is utilized to create a multilayer multi-\mno connectivity map. Each layer contains the measurements of one specific \kpi, which are aggregated geospatially based on a defined cell size $c$.
Fig.~\ref{fig:connectivityMap} shows an excerpt of the \ac{SINR} layer of the \acp{CM} for the different \mnos. The cell-wise determination of the best \mno finally leads to the multi-\mno map.

%
% Benefits
%
A straightforward estimation of achievable benefits by using the derived multi-\mno \ac{CM} for context-aware network selection is provided by Tab.~\ref{tab:multiMNO}, which shows the behavior of the network quality metrics \rsrp, \rsrq, \sinr and \cqi as well as of multiple end-to-end indicators. For each vehicle position within the real world traces, the statistical approach selects the \mno with the best network performance for the considered indicator.
%
% Tab. Multi-MNO
%
\newcommand{\cW}{0.6cm}

\aboverulesep = 1pt
\belowrulesep  = 1pt

\begin{table}[ht]
	\centering
	\caption{Network Quality Indicators and Active Measurements for Different Individual and Combined \mnos.}
	\begin{tabular}{p{1.35cm}|p{\cW}p{\cW}|p{\cW}p{\cW}|p{\cW}p{\cW}|p{\cW}}

		& \multicolumn{2}{c|}{\textbf{\mno A}} & \multicolumn{2}{c|}{\textbf{\mno B}} & \multicolumn{2}{c|}{\textbf{\mno C}} & \textbf{Multi}\\
		\textbf{Indicator} & \textbf{Mean} & \textbf{Best} & \textbf{Mean} & \textbf{Best} & \textbf{Mean} & \textbf{Best} & \textbf{\mno} \\

		\toprule

		\textbf{Coverage} & \textbf{1.0} & - & 0.96 & - & 0.91 & - & \textbf{1.0} \\
		\textbf{\rsrp~[dBm]} & \textbf{-87.6} & 0.525 & -97.1 & 0.19 & -91.4 & 0.285 & \textbf{-82.6} \\
		\textbf{\rsrq~[dB]} & \textbf{-7.5} & 0.684 & -9.85 & 0.2 & -10.7 & 0.116 & \textbf{-6.69} \\
		\textbf{\sinr~[dB]} & \textbf{14} & 0.556 & 7.23 & 0.229 & 7.43 & 0.216 & \textbf{18.2} \\
		\textbf{\cqi} & \textbf{10.1} & 0.491 & 9.25 & 0.317 & 8.32 & 0.191 & \textbf{12.2} \\
		
		\midrule
		$\textbf{\rtt~[ms]}$ & \textbf{57.4} & 0.652 & 380 & 0.159 & 459 & 0.189 & \textbf{43.6} \\
		\textbf{P}$\mathbf{_{TX}}$\textbf{~[dBm]} & \textbf{12.4} & 0.434 & 16.5 & 0.201 & 13.8 & 0.365 & \textbf{8.91} \\

	    \bottomrule		
		
	\end{tabular}
	
	\label{tab:multiMNO}
	\vspace{0.1cm}
	\emph{Best}: Proportion of the \mno providing the best metric value of all \mnos.

\end{table}

%
% Multi-MNO Benefits
%
Although the mean \rtt of \mnoB and \mnoC is much higher than the one of \mnoA, the multi-\mno is still able to improve the average behavior of all considered indicators. In addition, the mean \rtt is reduced by 24~\% and the average transmission power $\text{P}_\text{{TX}}$ is reduced by 28~\%.

\section{Client-based Online Data Rate Prediction} \label{sec:prediction}

%
% Fig. Correlation
%
\begin{figure*}[t] 
	\centering
	\includegraphics[width=\triple]{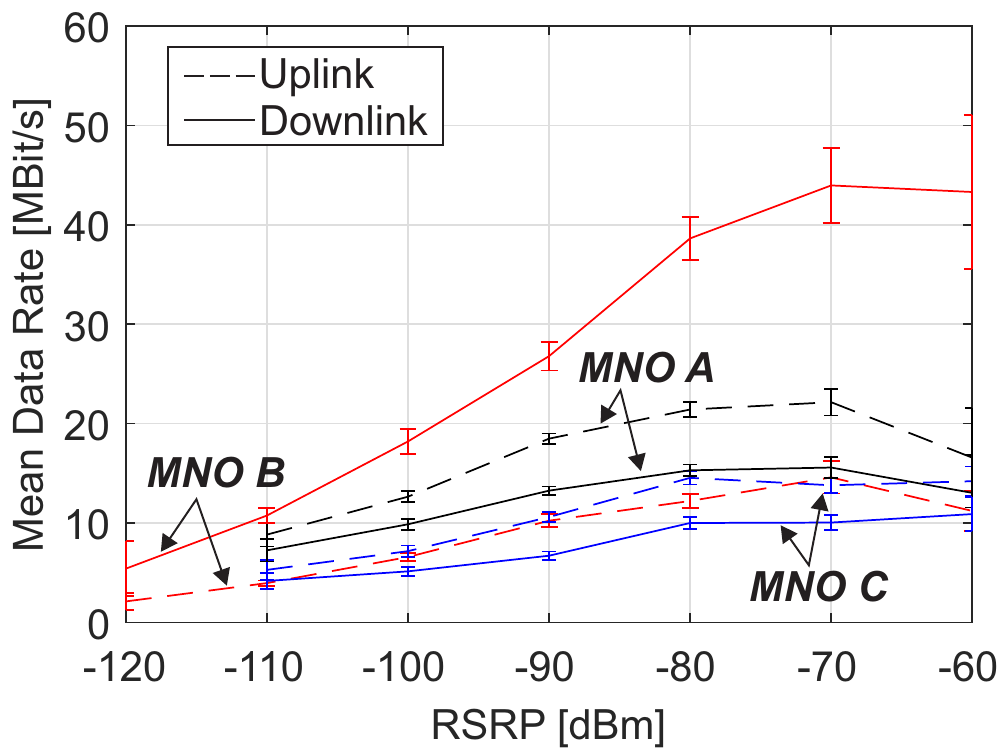}\quad
	\includegraphics[width=\triple]{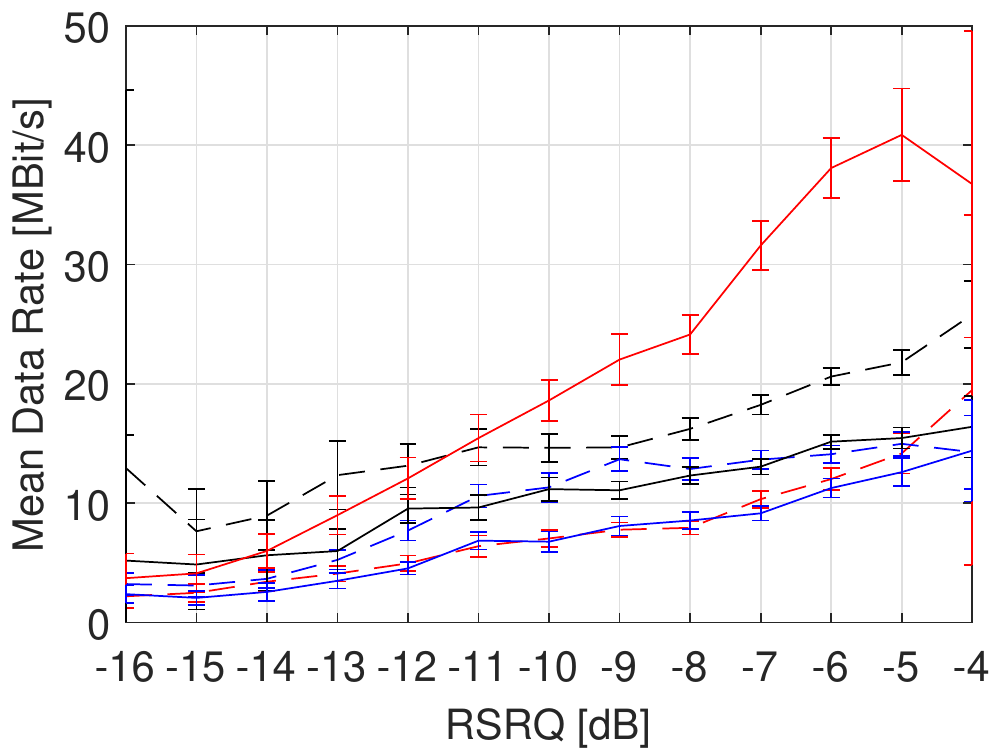}\quad
	\includegraphics[width=\triple]{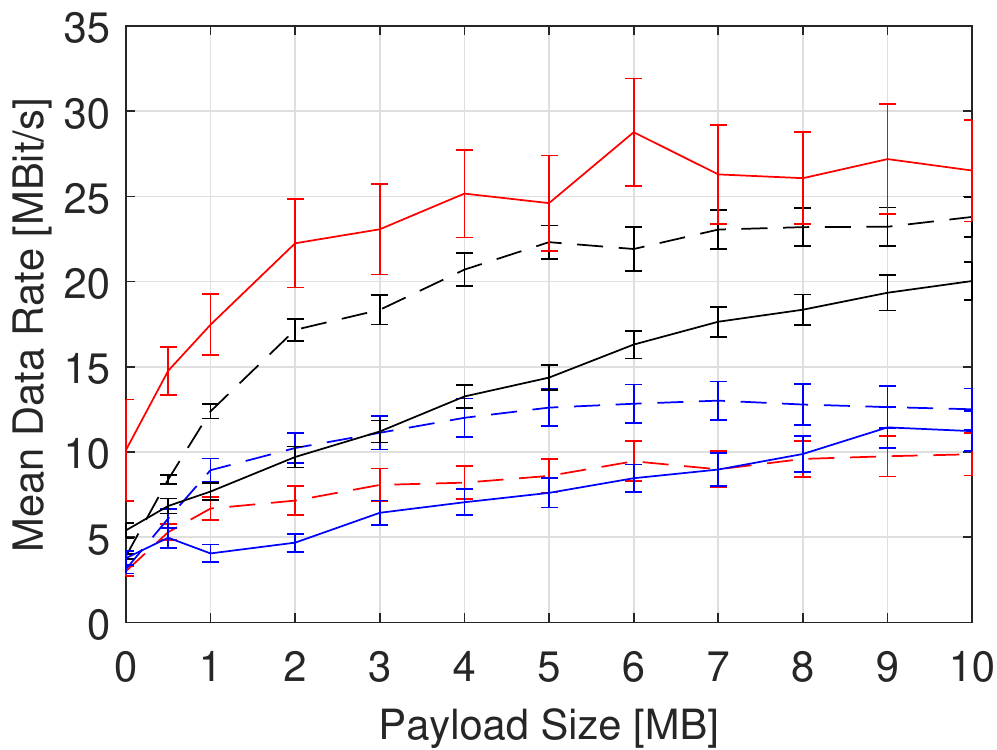}
	
	\includegraphics[width=\triple]{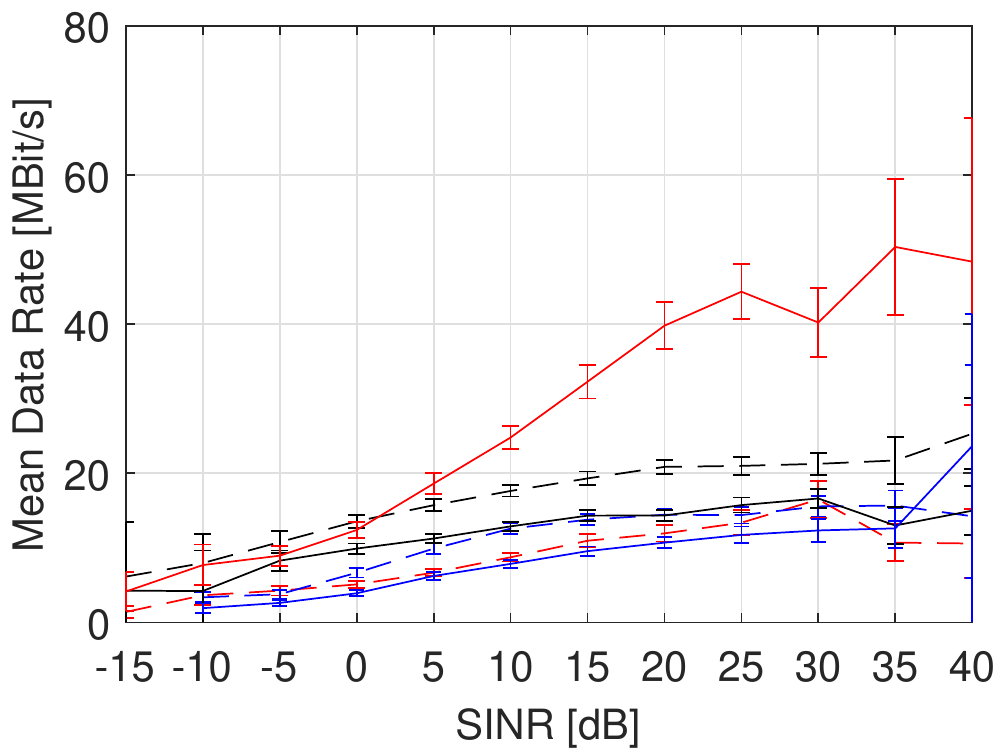}\quad
	\includegraphics[width=\triple]{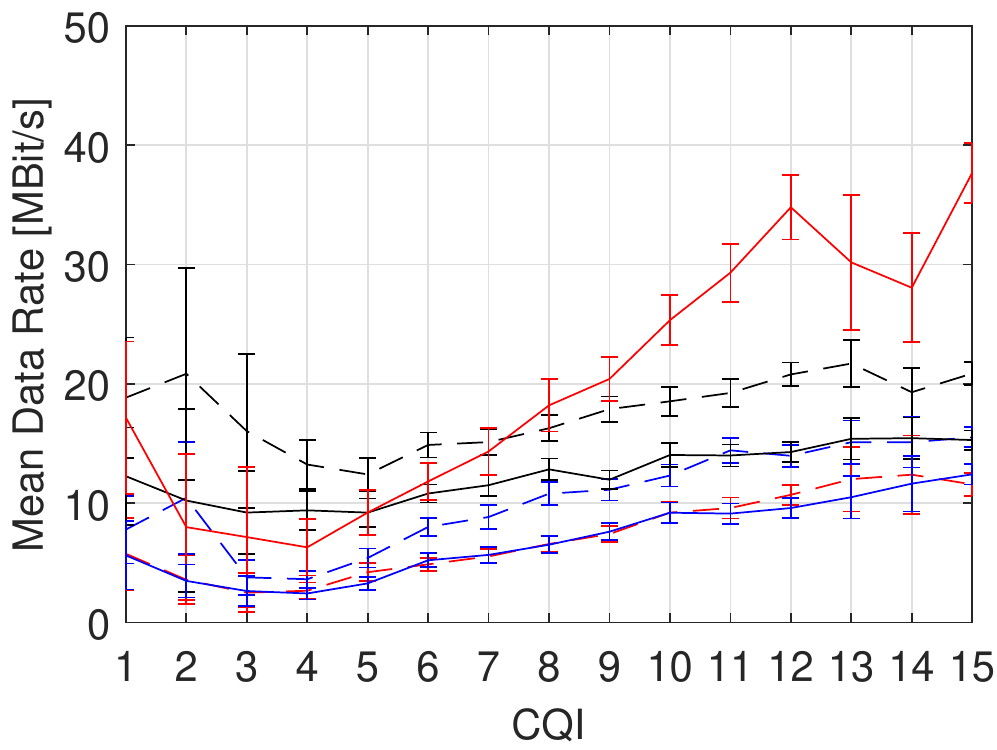}\quad
	\includegraphics[width=\triple]{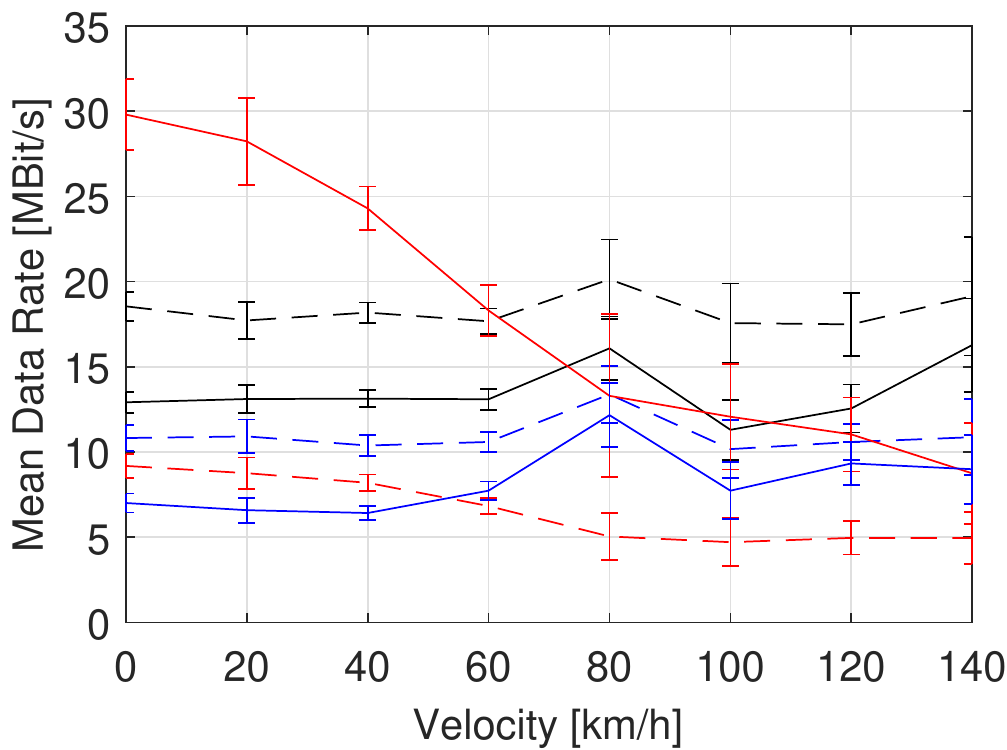}
	\vspace{-0.7cm}	
	\caption{Impact of the different indicators on the average resulting data rate for three network operators in uplink and downlink duration. The error bars illustrate the 0.95-confidence interval of the mean value.}
	\label{fig:correlation}
	\vspace{-0.3cm}	
\end{figure*}
%
%
%

%
% Introduction
%					
In this section, the results and insights of Sec.~\ref{sec:data} are exploited for predicting the end-to-end data rate. Furthermore, the importance of the individual features is analyzed and multi-\mno \acp{CM} are exploited to optimize the prediction accuracy.

%
% Models
%
Machine learning-based throughput prediction is a regression task. In the following, we analyze the performance of the \ann \cite{LeCun/etal/2015a}, \ac{M5} \cite{Quinlan/1992a}, \rf \cite{Breiman/2001a}, and \svm \cite{Cortes/Vapnik/1995a} models.
%
% Fig. Prediction Approach
%
\basicFig{}{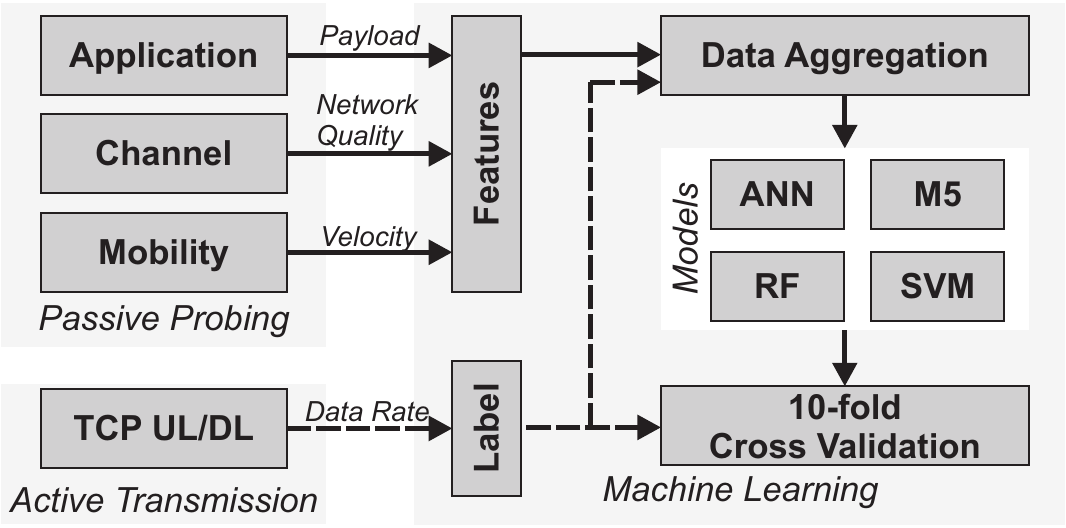}{Architecture model for client-based data rate prediction.}{fig:prediction_model}{-0.6cm}{0cm}{0.85}
%
% Features and Prediction Models
%
The prediction process is shown in Fig.\ref{fig:prediction_model}. During the \emph{training} phase, active data transmissions are performed and the resulting throughput and \rtt are captured and used as the \emph{labels} for the prediction. The actual training of the learning models is performed offline. The trained models are applied online during the \emph{application} phase.
%
% Models
%
The parameters of all considered models are optimized in a preprocessing step. For the \ann, a deep neural network with two hidden layers (10 and 5 neurons) resulted in the best performance. Additional parameters (learning rate $\eta=0.1$ , momentum $\alpha=0.001$) were tuned based on an evolutionary algorithm.  
The applied \rf consists of with 100 random trees and the \svm is parameterized as a linear L2/L2 \svm with squared hinge loss.
If not stated otherwise, all prediction results are obtained from 10-fold cross validation.

%
% Features
%
The applied feature vector consists of nine different features:
\begin{itemize}
	\item As \textbf{application context features}, the payload size of the data packet
	\item \textbf{Channel context features} contain \rsrp, \rsrq, \sinr, \cqi, \ac{TA} and the carrier frequency $f$
	\item As \textbf{mobility context features}, the current velocity of the vehicle and the cell id
\end{itemize}

%
% Performance Metrics
%
The \emph{coefficient of determinantion} ($R^{2}$) is a statistical metric for analyzing the prediction performance of a regression model, which describes the amount of the response variable variation that can be explained by derived model. It is widely used for data rate prediction (e.g., \cite{Jomrich/etal/2018a, Samba/etal/2017a}) and computed as 
%
% Eq. R^2
%
\begin{eqnarray}
	R^{2} = 1- \frac{\sum_{i=1}^{N}\left(\tilde{y}_{i} - y_{i} \right)^{2}}{\sum_{i=1}^{N}\left(\bar{y} - y_{i} \right)^{2}}
\end{eqnarray}
with $\bar{y}_{i}$ being the mean measurement value, $y_{i}$ being the current measurement and $\tilde{y}_{i}$ being the current prediction.

%
% Correlation 
%
Before the actual prediction results are discussed, the correlation between different indicators and the resulting data rate is analyzed in Fig.~\ref{fig:correlation}. 
%
% UL vs DL 
%
As it is also shown in Fig.~\ref{fig:kpi_histograms_highway}, \mnoB provides the highest data rates in the downlink direction.

%
% Indicators
%
The payload size has a significant impact on the achievable data rate for all \mnos and transmission directions. It is directly related to the slow start mechanism of \tcp and the channel coherence time. However, the upper bound of the mean data rate is lower than the one of the network quality indicators. Although this fact shows that the channel quality is an important indicator for the achievable data rate, the consideration of a single network quality indicator does not allow to derive meaningful conclusions for the resulting end-to-end data rate. Therefore, it can be expected that the machine learning-based joint consideration of multiple indicators is able to better describe the contained cross-dependencies.

%
% Speed
%
Apart from \mnoB, the velocity does not have a severe impact on the data rate. As discussed earlier (see Sec.~\ref{sec:data}), the average \ac{RSRP} for \mnoB is lower than for the other \mnos. With higher speeds, the probability for cellular handovers is increased, which severely disturb active data transmissions.

\subsection{Comparison of Different Data Aggregation Approaches}

In order to determine the optimal properties of the training sets, different data aggregation methods are analyzed. There is a trade-off between using a higher amount of training data (a single global data set) and a deeper consideration of the infrastructure-specific configurations (many local data sets). Therefore, the following measurement subsets are evaluated for all of the considered prediction models:
%
% Aggregation Levels
%
\begin{itemize}
	\item \textbf{\mno} uses the whole data per \mno (3 sets).
	\item \textbf{Scenario} divides the data for the different scenarios \emph{campus}, \emph{urban}, \emph{suburban} and \emph{highway} (12 sets).
	\item \textbf{\enb} groups the data based on the \enb id (105 sets).
	\item \textbf{Cell} performs cell id-specific data aggregation (220 sets).
\end{itemize}
%
% Tab. Throughput Prediction
%
\renewcommand{\cW}{0.8cm}

\aboverulesep = 1pt
\belowrulesep  = 1pt

\begin{table*}[ht]
	\centering
	\caption{Coefficient of Determination ($R^2$) for Different Machine Learning Models and Data Aggregation Levels}
	\begin{tabular}{p{0.2cm}p{1cm}|p{\cW}p{\cW}p{\cW}p{\cW}|p{\cW}p{\cW}p{\cW}p{\cW}|p{\cW}p{\cW}p{\cW}p{\cW}p{\cW}}

		& & \multicolumn{4}{c|}{\textbf{\mno A}} & \multicolumn{4}{c|}{\textbf{\mno B}} & \multicolumn{4}{c}{\textbf{\mno C}} \\

		& \textbf{Data} & \textbf{\ann} & \textbf{M5} &\textbf{\ac{RF}} & \textbf{\svm} & \textbf{\ann} & \textbf{M5} &\textbf{\ac{RF}} & \textbf{\svm} & \textbf{\ann} & \textbf{M5} &\textbf{\ac{RF}} & \textbf{\svm} \\
		\toprule	
		\sideHeader{5}{0.5cm}{Uplink} 
		& \textbf{\mno} & 0.685 & 0.754 & \textbf{0.8} & 0.71 & 0.46 & 0.658 & \textbf{0.707} & 0.594 & 0.69 & 0.779 & \textbf{0.82} & 0.728 \\
		& \textbf{Scenario} & 0.729 & 0.779 & \textbf{0.806} & 0.683 & 0.49 & 0.572 & \textbf{0.633} & 0.555 & 0.489 & 0.64 & \textbf{0.686} & 0.572 \\
		& \textbf{\enb} & 0.578 & 0.724 & \textbf{0.731} & 0.592 & 0.285 & 0.432 & \textbf{0.456} & 0.44 & 0.384 & 0.57 & \textbf{0.604} & 0.512 \\
		& \textbf{Cell} & 0.532 & 0.687 & \textbf{0.715} & 0.58 & 0.275 & 0.412 & \textbf{0.444} & 0.397 & 0.355 & \textbf{0.505} & \textbf{0.505} & 0.424 \\

		\midrule
		\sideHeader{6}{0.5cm}{Downlink} 
		& \textbf{\mno} & 0.499 & 0.603 & 0.591 & \textbf{0.612} & 0.524 & 0.584 & \textbf{0.648} & 0.578 & 0.41 & 0.504 & \textbf{0.552} & 0.531 \\
		& \textbf{Scenario} & 0.551 & 0.62 & 0.615 & \textbf{0.627} & 0.321 & 0.491 & \textbf{0.541} & 0.496 & 0.265 & 0.386 & \textbf{0.422} & 0.41 \\
		& \textbf{\enb} & 0.34 & 0.551 & 0.552 & \textbf{0.58} & 0.263 & 0.317 & 0.357 & \textbf{0.362} & 0.151 & 0.323 & 0.334 & \textbf{0.361} \\
		& \textbf{Cell} & 0.3 & \textbf{0.564} & 0.503 & 0.555 & 0.258 & 0.325 & \textbf{0.379} & 0.372 & 0.19 & 0.296 & \textbf{0.306} & 0.294 \\
	    \bottomrule		
		
	\end{tabular}
	
	\vspace{0.1cm}
	\emph{ANN}: Artificial Neural Network, \emph{M5}: M5 Regression Tree, \emph{RF}: Random Forest, \emph{SVM}: Support Vector Machine 
	\label{tab:throughput}
	\vspace{-0.3cm}
\end{table*}

%
% Fig. Model Performance
%
\newcommand{\sfw}{0.32}
\begin{figure*}[t] 
	\centering
	
	\subfig{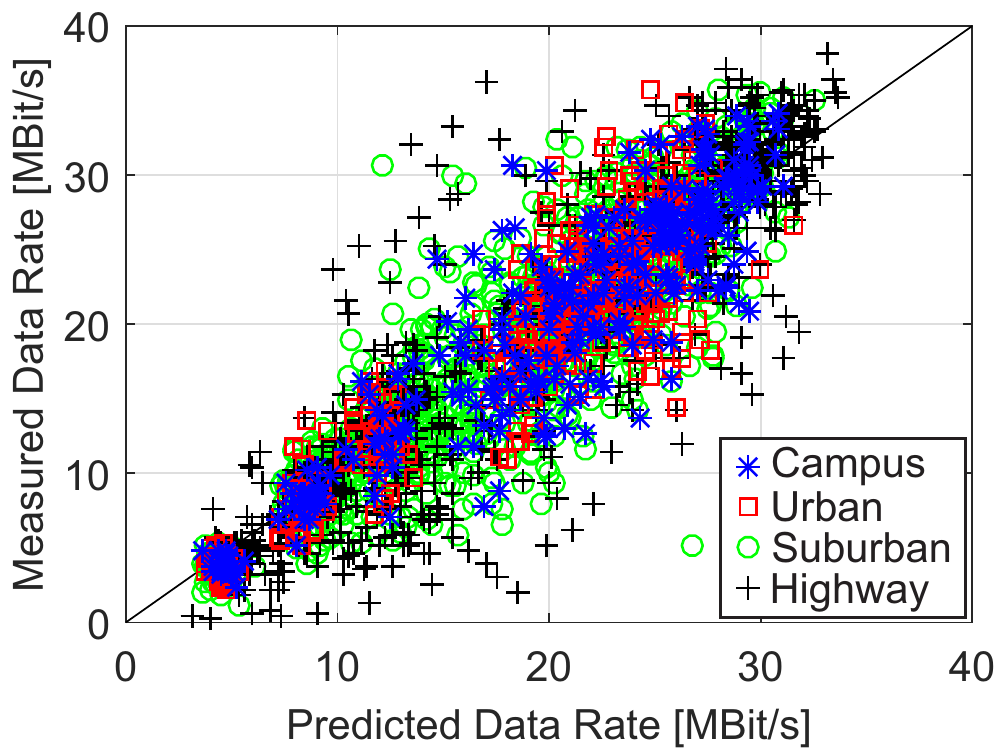}{\sfw}{MNO~A}
	\subfig{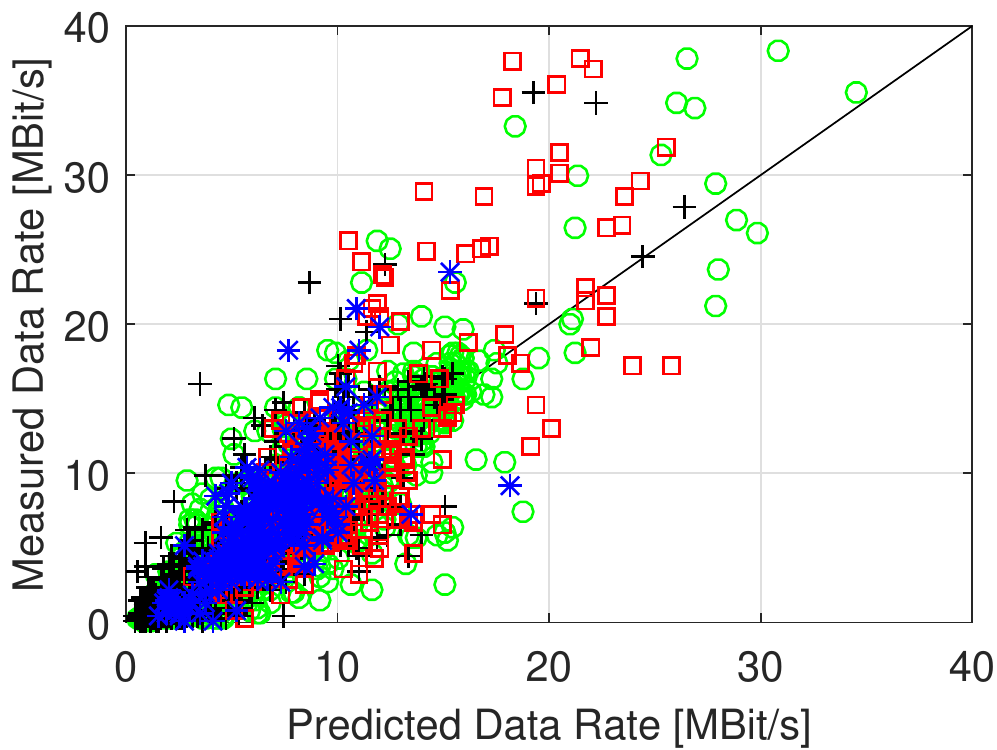}{\sfw}{MNO~B}
	\subfig{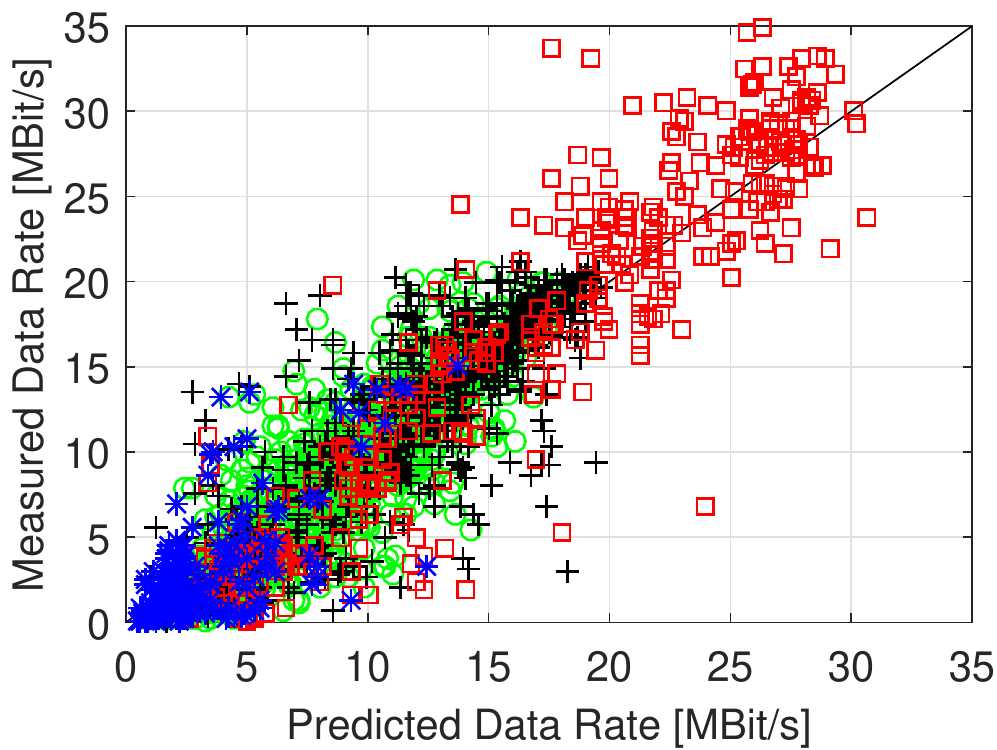}{\sfw}{MNO~C}
	
	\caption{Comparison of predictions and measurements for \rf-based uplink data rate prediction for the considered \mnos in different scenarios. The black diagonal line illustrates perfect prediction. Data points in the upper left triangle represent underestimations of the data rate, elements in the lower right triangle overestimate the data rate.}
	\label{fig:model_performance}
	\vspace{-0.5cm}
\end{figure*}
Tab.~\ref{tab:throughput} summarizes the results of the data rate prediction in uplink and downlink direction.
%
% Global vs Local
%
For all considered \mnos, it can be seen that the $R^2$ is reduced when the grade of locality is increased. The highest value is achieved by the global data set, which contains all measurements per \mno. \keyResult{The prediction models benefit more from the additional data than from more context-specific data sets.}

%
% Random Forest 
%
In most cases, the best prediction performance is achieved with the \rf model. A possible explanation for this behavior is that in many situations, one of the considered network quality indicators has a dominant impact on the behavior of the data rate under defined conditions. 
The analysis in \cite{Sliwa/etal/2018b} points out that at the cell edge -- which can be identified by the \ac{RSRP} -- the interference level, which is partly identifiable by the \ac{RSRQ}, has a strong impact on the data rate. In contrast to that, the \ac{SINR} is of higher importance within the cell center. These interval-wise scope regions match well with the general tree-like structure of \rf and \ac{M5}.
Since \ac{CART}-based models can be implemented in a very resource-efficient way as a sequence of \texttt{if/else} statements, this fact has positive implications on the resource efficiency for the online data rate prediction in the application phase of the models.
%
% M5
%
\keyResult{In many cases, the \ac{M5} approach achieves the second highest $R^2$ value. This is remarkable, as the resulting trained model is significantly smaller than the \rf. As an example for the full uplink data set of \mnoA, the \rf consists of 120533 leafs (numerical values) but the \ac{M5} has only 11 leafs (linear regression models).} As a consequence, the \ac{M5} could be applied as a lightweight alternative to the \rf for memory-constrained systems (e.g., microcontrollers).
%
% MNO
%
It can be seen that the highest $R^2$ values are achieved for \mnoA and \mnoC. As the average \rsrp of \mnoB is low (see Tab.~\ref{tab:multiMNO}), the probability for cellular handovers, which decrease the prediction accuracy, is higher than for the other \mnos, especially in the \emph{campus} and \emph{highway} scenarios. 

%
% Uplink vs Downlink
%
\keyResult{The prediction works more precise in the uplink direction.} As the traffic load in the uplink is usually much lower than in the downlink direction \cite{Bui/etal/2017a}, the \ue less likely encounters a loaded cell. Therefore, it can be concluded that the data rate is more impacted by the channel quality-related effects -- which are covered to a high grade by the applied feature set -- than by resource competition and scheduling-related aspects. While the consideration of the latter could likely improve the download prediction accuracy, this approach would require infrastructure side knowledge, which cannot be accessed using the proposed client-based approach.
%
% Conclusion
%
Based on the obtained insights, the following evaluations focus on \rf-based uplink data rate prediction using the global \mno data sets.

%
% Fig. RF Uplink Prediction
%
A graphical representation of the resulting uplink data rate prediction performance of the trained \rf models is shown in Fig.~\ref{fig:model_performance}. It can be seen that the value range, error spread and the scenario-dependency is different for the considered \mnos. While \mnoA shows a homogeneous behavior for all scenarios, both other \mnos have defined focus regions for the provided coverage.
%
% Fig. Confusion Matrices
%
\renewcommand{\sfw}{0.24}
\begin{figure*}[] 
	\centering
	
	\subfig{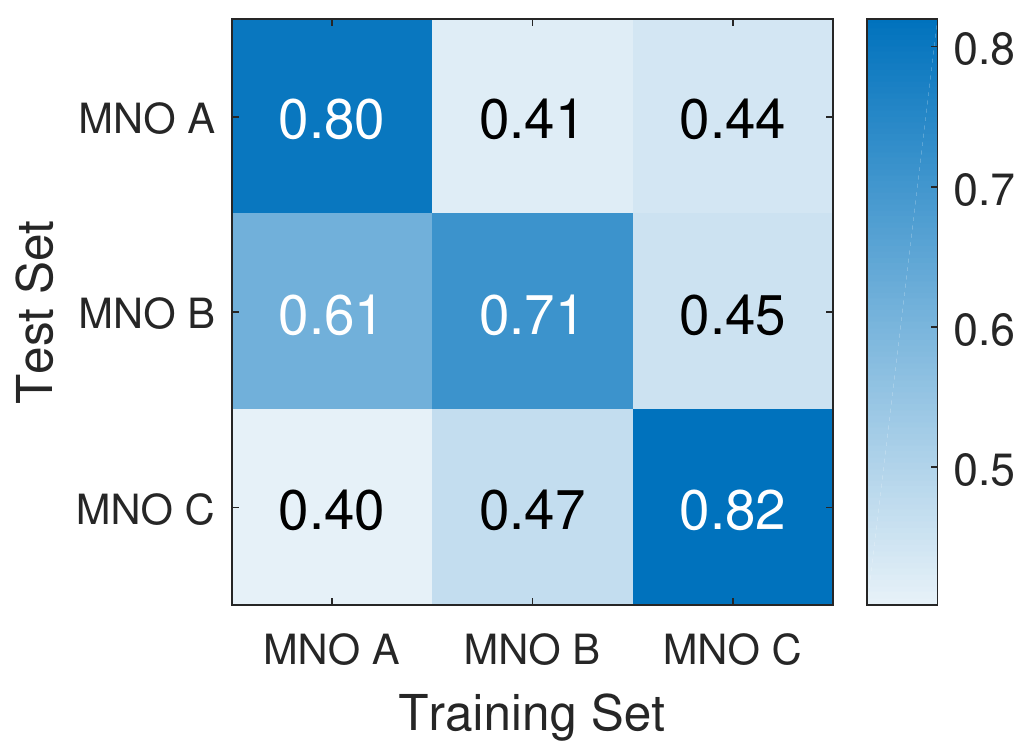}{\sfw}{Cross-\mno}
	\subfig{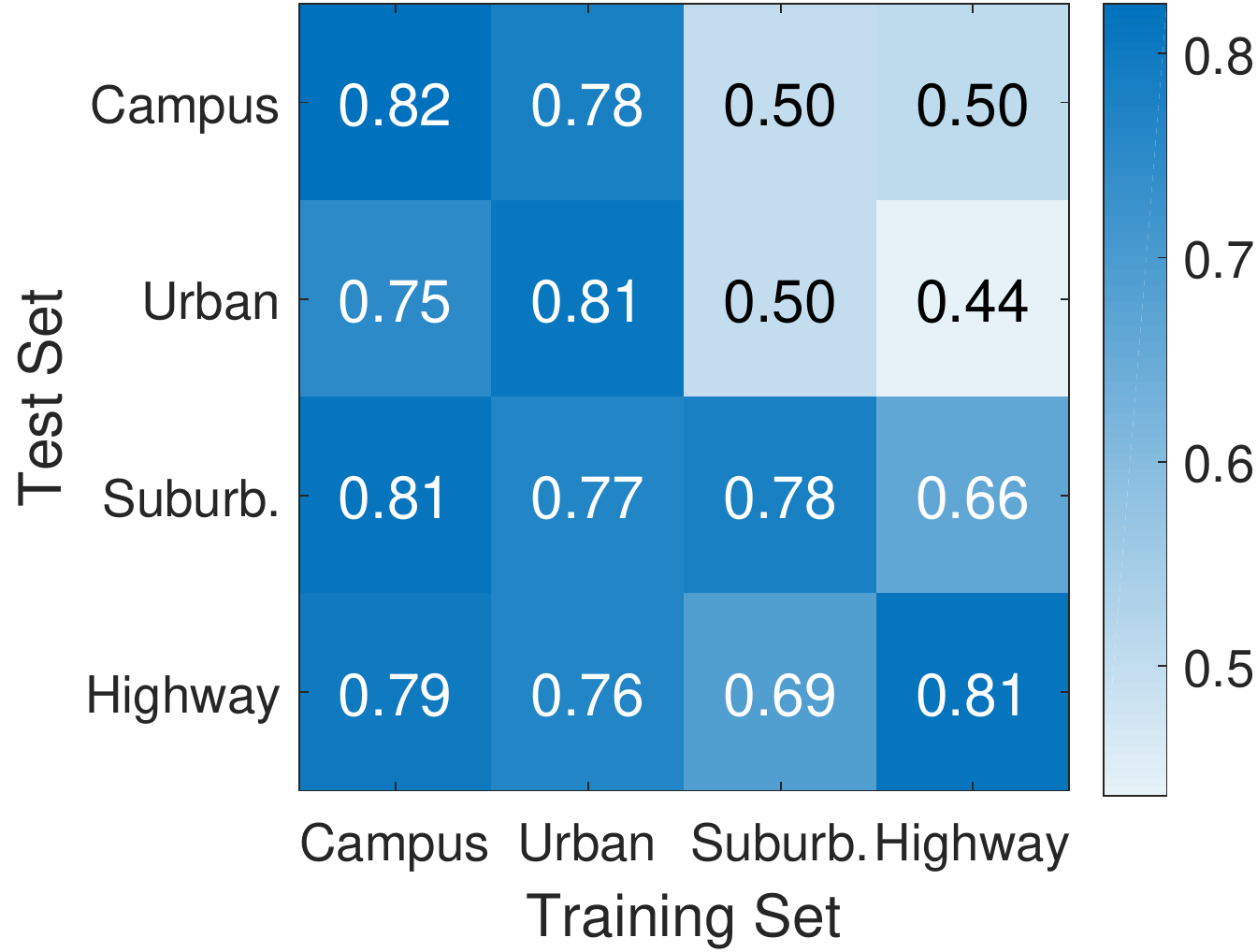}{\sfw}{\mno A}
	\subfig{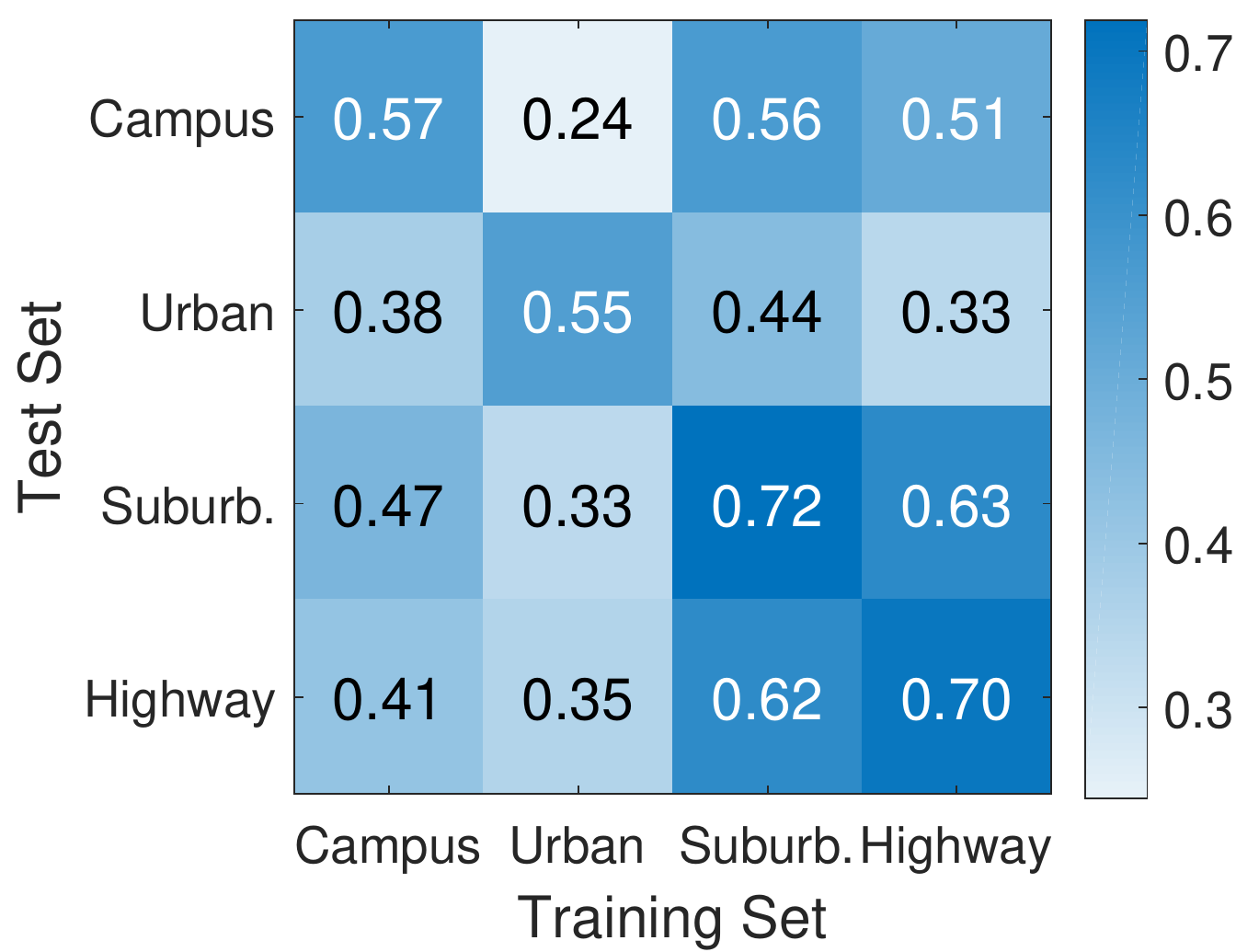}{\sfw}{\mno B}
	\subfig{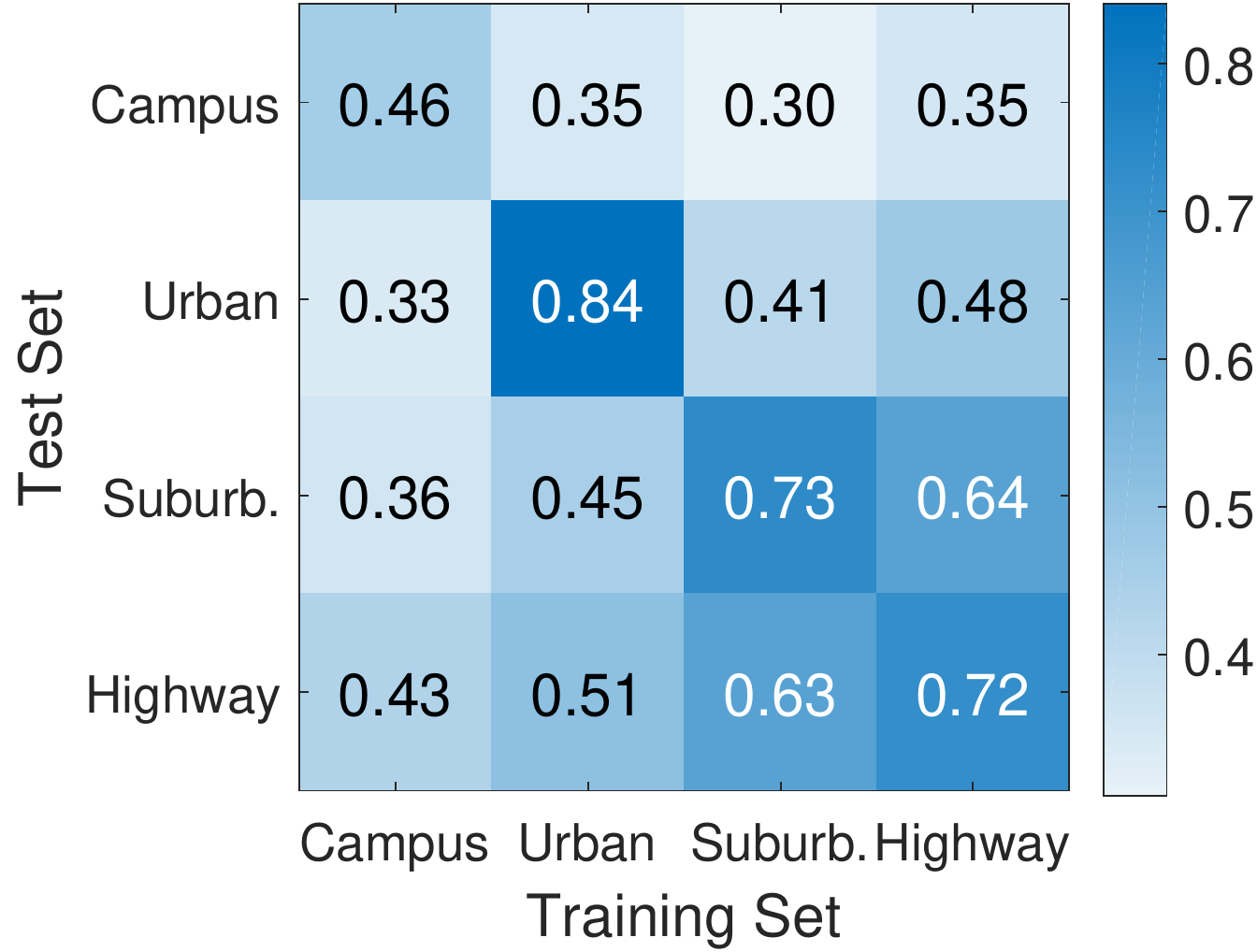}{\sfw}{\mno C}
	
	\caption{$R^2$ results for \rf-based uplink data rate prediction with different combinations of training and test sets. \emph{Note}: For each element of the main diagonal, the results show the 10-fold cross validation on the training data in order to avoid interdependencies between training set and test set.}
	\label{fig:confusion}
	\vspace{-0.5cm}
\end{figure*}
Fig.~\ref{fig:confusion} shows the results of different combinations of training and test sets.
%
% Cross MNO
%
The cross-\mno applicability of a trained model is evaluated in Fig.~\ref{fig:confusion}~(a). It can be seen that\keyResult{multi-\mno data rate prediction requires \mno-specific training of the machine learning models.} This behavior can be explained by the usage of different network configurations with respect to resource scheduling, applied transmission power and distribution of the carrier frequencies.

%
% Cross Scenario per MNO
%
Furthermore, the applicability of cross-scenario training and testing is shown in Fig.~\ref{fig:confusion}~(b)-(d). For each \mno, a model is trained on the data of each individual scenario and tested against the data of the other scenarios.
\mnoA achieves very good generalization for the campus and urban scenarios but not for the suburban and highway scenarios. The results can be better understood by considering Fig.~\ref{fig:model_performance}(a). The prediction model is very homogeneous for \mnoA but the highest variance is shown by data points from the highway and suburban data set. 
% stop and go traffic not contained in suburb + highway?
%
% MNO C Campus
%
For \mnoC, the \lte coverage in the campus scenarios is only 76.25~\%, which leads to a bad prediction performance for this scenario.

\subsection{Feature Importance}

%
% Fig. Feature Importance
%
\begin{figure}[]
	\centering		  
	\includegraphics[width=1\columnwidth]{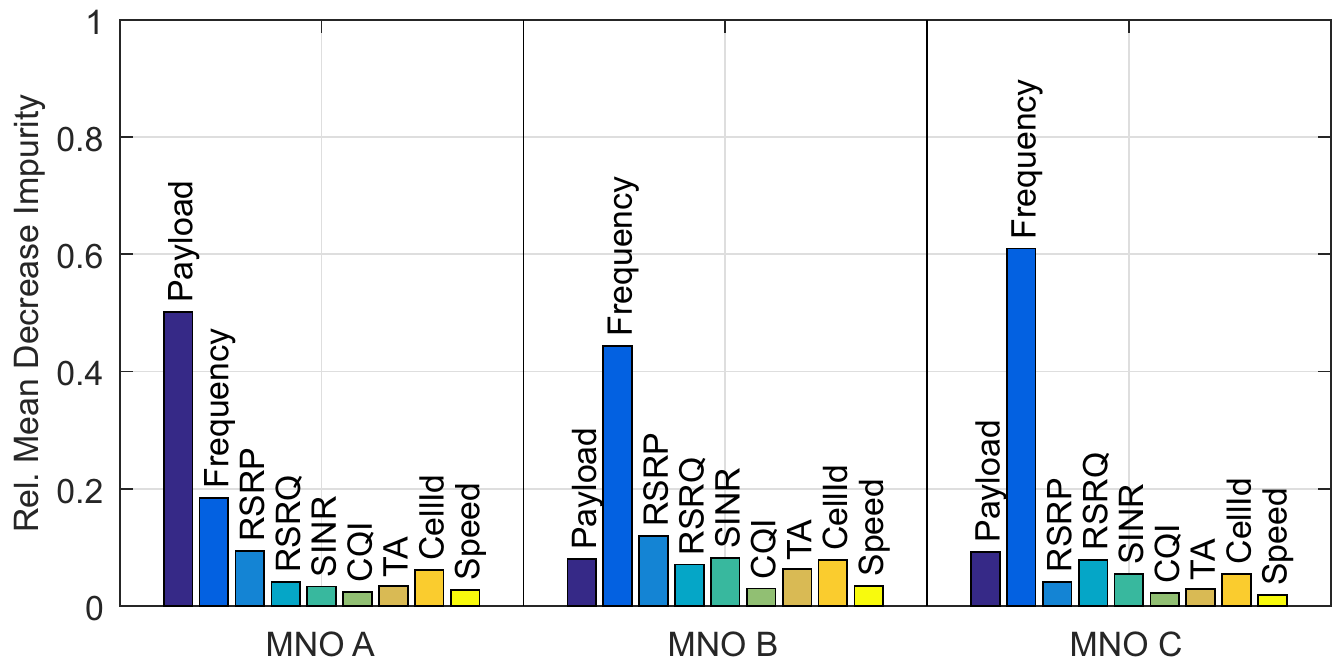}
	\vspace{-0.5cm}	
	\caption{Importance of individual features for \rf-based uplink data rate prediction.}
	\label{fig:featureImportance}
	\vspace{-0.5cm}	
\end{figure}
Fig.~\ref{fig:featureImportance} shows the relative average entropy-based \ac{MDI} \cite{Louppe/etal/2013a} for the \rf data rate prediction model for each \mno. Again, the considered \mnos show different behaviors. 
%
% Comparison to Related Work - Jomrich/Samba
%
The features importance explains why the achieved overall prediction accuracy is significantly higher than the results achieved by related work, which only rely on passively measured network quality indicators (\cite{Jomrich/etal/2018a, Samba/etal/2017a, Riihijarvi/Mahonen/2018a}). 
%
% Importance of the Payload Size
%
In addition to the carrier frequency, the payload size has a dominant impact on the prediction accuracy as it has different interdependencies to the network quality and the applied protocols. Larger packets likely correspond to longer transmission durations, which means that the dependency to the channel coherence time is increased as the vehicle is moving during the transmission process. Moreover, the \tcp slow start is severely dependent on the packet size.
%
% Irrelevant features: RTT, Time of Day
%
For completeness, it should be remarked that the impact of other indicators (\rtt, time of day) was analyzed in a preparatory step. As the integration of these indicators did not result in better prediction performance, they were not further evaluated. We conclude that the added information is already immanently represented by the considered features, e.g. time of day has an impact on the cell load, which is correlated to the \rsrq.

\subsection{Exploiting Multi-\mno Connectivity Maps for Data Rate Prediction}

In the following, we analyze the suitability of using \acp{CM} for uplink data rate prediction in vehicular networks. In contrast to the previous measurement-based predictions approaches, the network quality is now solely looked up from the a priori information of the \acp{CM} based on the vehicle's location without performing dedicated radio quality measurements. 
%
% Tab. Connectivity Maps
%
%\newcommand{\cW}{0.8cm}

\aboverulesep = 1pt
\belowrulesep  = 1pt

\begin{table}[ht]
	\centering
	\caption{$R^2$ and \mae for \rf-based Prediction with Connectivity Maps and Different Cell Sizes in the Uplink}
	\begin{tabular}{p{1cm}|p{\cW}p{\cW}|p{\cW}p{\cW}|p{\cW}p{\cW}}

		& \multicolumn{2}{c|}{\textbf{\mno A}} & \multicolumn{2}{c|}{\textbf{\mno B}} & \multicolumn{2}{c}{\textbf{\mno C}}\\

		\textbf{Data} & \textbf{$\mathbf{R}^2$} & \textbf{\mae} & \textbf{$\mathbf{R}^2$} & \textbf{\mae} & \textbf{$\mathbf{R}^2$} & \textbf{\mae} \\
		\toprule

		\textbf{\mno} & 0.8 & 2.92 & 0.707 & 2.3 & 0.82 & 2.27 \\
		$\mathbf{CM_{5}}$ & 0.797 & 3.02 & 0.71 & 2.23 & 0.851 & 2.16 \\
		$\mathbf{CM_{10}}$ & \textbf{0.861} & \textbf{2.42} & 0.716 & 2.22 & 0.852 & 2.15 \\
		$\mathbf{CM_{25}}$ & 0.856 & 2.56 & \textbf{0.719} & \textbf{2.21} & \textbf{0.852} & \textbf{2.11} \\
		$\mathbf{CM_{50}}$ & 0.857 & 2.43 & 0.719 & 2.22 & 0.851 & 2.12 \\
		$\mathbf{CM_{100}}$ & 0.844 & 2.54 & 0.714 & 2.22 & 0.846 & 2.14 \\

	    \bottomrule		
		
	\end{tabular}

	\vspace{0.1cm}
	$\textit{CM}_{c}$: Connectivity map with cell size $c$ [m], \emph{\mno}: Total operator data set
	\label{tab:connectivityMaps}

\end{table}

Tab.~\ref{tab:connectivityMaps} summarizes the $R^2$ and \mae values for the considered \mnos using different \ac{CM} cell sizes. As a reference, the results for measurement-based \mno without \acp{CM} are shown.
%
% Cellwith
%
The choice of the cell size value represents the trade-off between using more data for the learning model and modeling a more specific representation of the environment. It can be assumed that the location information is influenced by \gps positioning errors. Furthermore, the cell size defines the storage complexity of the \ac{CM} model.
It can be seen that the \ac{CM}-based approach has different implications for the considered \mnos. While \emph{\mno B} does not significantly benefit from using \acp{CM}, \emph{\mno A} and \emph{\mno C} achieve significant improvements of the prediction accuracy (for \mnoA, the \mae is reduced by 17~\% / 0.5~MBit/s).
The prediction benefits from the cell-wise data aggregation performed by the \acp{CM}, which is able to compensate short-term influences such as multi-path fading. As a result, overfitting to a very specific radio channel situation is avoided and the resulting model achieves a better generalization.

For completeness, it should be noted that we investigated the combined usage of \ac{CM} as well as measured features. Since this approach resulted in a similar performance as the pure \ac{CM}-based version, it is not further evaluated. 

\section{Conclusion}

%
% Introduction
%
In this paper, we presented an empirical evaluation of multi-\mno communication for vehicular networks and analyzed the performance of multiple models for online data rate prediction. 
%
% Benefits of MC
%
The analysis shows that even if one of the \mnos provides superior coverage and network quality, the overall performance can still be increased significantly by applying a multi-\mno approach. 
%
% MNO-specific Prediction Models
%
The achievable accuracy of data rate prediction is highly \mno-dependent. Therefore, data-driven multi-\mno approaches require different learning models for each considered \mno, which consequently increases the importance of the resource efficiency of the prediction models. In the majority of the considered cases, the \rf model achieved the best prediction accuracy.

%
% Payload Size
%
Although the channel properties have a significant impact on the achievable data rate, they need to be set into relation to the payload size in order to achieve precise predictions. This way, the machine learning process is able to consider hidden interdependencies between channel coherence time and network protocols.
%
% Benefits of Connectivity Maps
%
Connectivity maps can be utilized as a priori information to compensate short-term influences and can help to achieve a better generalization of the prediction schemes by avoiding overfitting. However, the achievable benefits of this approach and the optimal parametrization are also \mno-specific.
%
% Future work
%
In future work, we will leverage the prediction models for optimizing communication processes within context-predictive vehicular networks.

\section*{Acknowledgment}

\footnotesize
Part of the work on this paper has been supported by Deutsche Forschungsgemeinschaft (DFG) within the Collaborative Research Center SFB 876 ``Providing Information by Resource-Constrained Analysis'', project B4.

\bibliographystyle{IEEEtran}
\bibliography{Bibliography}

\begin{thebibliography}{10}
\providecommand{\url}[1]{#1}
\csname url@samestyle\endcsname
\providecommand{\newblock}{\relax}
\providecommand{\bibinfo}[2]{#2}
\providecommand{\BIBentrySTDinterwordspacing}{\spaceskip=0pt\relax}
\providecommand{\BIBentryALTinterwordstretchfactor}{4}
\providecommand{\BIBentryALTinterwordspacing}{\spaceskip=\fontdimen2\font plus
\BIBentryALTinterwordstretchfactor\fontdimen3\font minus
  \fontdimen4\font\relax}
\providecommand{\BIBforeignlanguage}[2]{{%
\expandafter\ifx\csname l@#1\endcsname\relax
\typeout{** WARNING: IEEEtran.bst: No hyphenation pattern has been}%
\typeout{** loaded for the language `#1'. Using the pattern for}%
\typeout{** the default language instead.}%
\else
\language=\csname l@#1\endcsname
\fi
#2}}
\providecommand{\BIBdecl}{\relax}
\BIBdecl

\bibitem{Bui/etal/2017a}
N.~Bui, M.~Cesana, S.~A. Hosseini, Q.~Liao, I.~Malanchini, and J.~Widmer, ``A
  survey of anticipatory mobile networking: Context-based classification,
  prediction methodologies, and optimization techniques,'' \emph{IEEE
  Communications Surveys \& Tutorials}, 2017.

\bibitem{Sliwa/etal/2019b}
B.~Sliwa, T.~Liebig, T.~Vranken, M.~Schreckenberg, and C.~Wietfeld,
  ``System-of-systems modeling, analysis and optimization of hybrid vehicular
  traffic,'' in \emph{2019 Annual IEEE International Systems Conference
  (SysCon)}, Orlando, Florida, USA, Apr 2019.

\bibitem{Sliwa/etal/2018b}
B.~Sliwa, T.~Liebig, R.~Falkenberg, J.~Pillmann, and C.~Wietfeld, ``Efficient
  machine-type communication using multi-metric context-awareness for cars used
  as mobile sensors in upcoming {5G} networks,'' in \emph{2018 IEEE 87th
  Vehicular Technology Conference (VTC-Spring)}, Porto, Portugal, Jun 2018,
  {Best Student Paper Award}.

\bibitem{Sliwa/etal/2016b}
B.~Sliwa, D.~Behnke, C.~Ide, and C.~Wietfeld, ``{B.A.T.Mobile}: Leveraging
  mobility control knowledge for efficient routing in mobile robotic
  networks,'' in \emph{IEEE GLOBECOM 2016 Workshop on Wireless Networking,
  Control and Positioning of Unmanned Autonomous Vehicles (Wi-UAV)}, Washington
  D.C., USA, Dec 2016.

\bibitem{Sliwa/etal/2018a}
B.~Sliwa, T.~Liebig, R.~Falkenberg, J.~Pillmann, and C.~Wietfeld, ``Machine
  learning based context-predictive car-to-cloud communication using
  multi-layer connectivity maps for upcoming {5G} networks,'' in \emph{2018
  IEEE 88th Vehicular Technology Conference (VTC-Fall)}, Chicago, USA, Aug
  2018.

\bibitem{Sliwa/etal/2019d}
B.~Sliwa, R.~Falkenberg, T.~Liebig, N.~Piatkowski, and C.~Wietfeld, ``Boosting
  vehicle-to-cloud communication by machine learning-enabled context
  prediction,'' \emph{IEEE Transactions on Intelligent Transportation Systems},
  2019, {A}ccepted for publication.

\bibitem{Sliwa/Wietfeld/2019c}
B.~Sliwa and C.~Wietfeld, ``Towards data-driven simulation of end-to-end
  network performance indicators,'' in \emph{2019 IEEE 90th Vehicular
  Technology Conference (VTC-Fall)}, Honolulu, Hawaii, USA, Sep 2019.

\bibitem{Ye/etal/2018a}
H.~Ye, L.~Liang, G.~Y. Li, J.~Kim, L.~Lu, and M.~Wu, ``Machine learning for
  vehicular networks: {R}ecent advances and application examples,'' \emph{IEEE
  Vehicular Technology Magazine}, vol.~13, no.~2, pp. 94--101, June 2018.

\bibitem{Jiang/etal/2017a}
C.~Jiang, H.~Zhang, Y.~Ren, Z.~Han, K.~C. Chen, and L.~Hanzo, ``Machine
  learning paradigms for next-generation wireless networks,'' \emph{IEEE
  Wireless Communications}, vol.~24, no.~2, pp. 98--105, April 2017.

\bibitem{Walelgne/etal/2018a}
E.~A. Walelgne, J.~Manner, V.~Bajpai, and J.~Ott, ``Analyzing throughput and
  stability in cellular networks,'' in \emph{NOMS 2018 - 2018 IEEE/IFIP Network
  Operations and Management Symposium}, April 2018, pp. 1--9.

\bibitem{Akselrod/etal/2017a}
M.~Akselrod, N.~Becker, M.~Fidler, and R.~Luebben, ``{4G} {LTE} on the road -
  what impacts download speeds most?'' in \emph{2017 IEEE 86th Vehicular
  Technology Conference (VTC-Fall)}, Sep. 2017, pp. 1--6.

\bibitem{Cavalcanti/etal/2018a}
E.~R. Cavalcanti, J.~A.~R. de~Souza, M.~A. Spohn, R.~C. d.~M. Gomes, and A.~F.
  B. F.~d. Costa, ``{VANETs}' research over the past decade: {O}verview,
  credibility, and trends,'' \emph{SIGCOMM Comput. Commun. Rev.}, vol.~48,
  no.~2, pp. 31--39, May 2018.

\bibitem{Chaudhari/Biradar/2015a}
S.~S. Chaudhari and R.~C. Biradar, ``Survey of bandwidth estimation techniques
  in communication networks,'' \emph{Wireless Personal Communications},
  vol.~83, no.~2, pp. 1425--1476, Jul 2015.

\bibitem{Riihijarvi/Mahonen/2018a}
J.~Riihijarvi and P.~Mahonen, ``Machine learning for performance prediction in
  mobile cellular networks,'' \emph{IEEE Computational Intelligence Magazine},
  vol.~13, no.~1, pp. 51--60, Feb 2018.

\bibitem{Cainey/etal/2014a}
J.~Cainey, B.~Gill, S.~Johnston, J.~Robinson, and S.~Westwood, ``Modelling
  download throughput of {LTE} networks,'' in \emph{39th Annual IEEE Conference
  on Local Computer Networks Workshops}, Sep. 2014, pp. 623--628.

\bibitem{Jomrich/etal/2018a}
F.~Jomrich, A.~Herzberger, T.~Meuser, B.~Richerzhagen, R.~Steinmetz, and
  C.~Wille, ``Cellular bandwidth prediction for highly automated driving -
  {E}valuation of machine learning approaches based on real-world data,'' in
  \emph{Proceedings of the 4th International Conference on Vehicle Technology
  and Intelligent Transport Systems 2018}, no.~4.\hskip 1em plus 0.5em minus
  0.4em\relax SCITEPRESS, Mar 2018, pp. 121--131.

\bibitem{Samba/etal/2017a}
A.~Samba, Y.~Busnel, A.~Blanc, P.~Dooze, and G.~Simon, ``Instantaneous
  throughput prediction in cellular networks: {W}hich information is needed?''
  in \emph{2017 IFIP/IEEE Symposium on Integrated Network and Service
  Management (IM)}, May 2017, pp. 624--627.

\bibitem{Kousias/etal/2017a}
K.~Kousias, C.~Midoglu, O.~Alay, A.~Lutu, A.~Argyriou, and M.~Riegler, ``The
  same, only different: {C}ontrasting mobile operator behavior from
  crowdsourced dataset,'' in \emph{2017 IEEE 28th Annual International
  Symposium on Personal, Indoor, and Mobile Radio Communications (PIMRC)}, Oct
  2017, pp. 1--6.

\bibitem{Nikolov/etal/2018a}
G.~Nikolov, M.~Kuhn, and B.~Wenning, ``{UE}-based estimation of available
  uplink data rates in cellular networks,'' in \emph{2018 14th International
  Conference on Wireless and Mobile Computing, Networking and Communications
  (WiMob)}, Oct 2018, pp. 169--174.

\bibitem{Kasparick/etal/2016a}
M.~Kasparick, R.~L.~G. Cavalcante, S.~Valentin, S.~Stańczak, and M.~Yukawa,
  ``Kernel-based adaptive online reconstruction of coverage maps with side
  information,'' \emph{IEEE Transactions on Vehicular Technology}, vol.~65,
  no.~7, pp. 5461--5473, July 2016.

\bibitem{Apajalahti/etal/2018a}
K.~Apajalahti, E.~A. Walelgne, J.~Manner, and E.~Hyvönen, ``Correlation-based
  feature mapping of crowdsourced {LTE} data,'' in \emph{2018 IEEE 29th Annual
  International Symposium on Personal, Indoor and Mobile Radio Communications
  (PIMRC)}, Sep. 2018, pp. 1--7.

\bibitem{Sliwa/2019a}
\BIBentryALTinterwordspacing
B.~Sliwa, ``Raw experimental cellular network quality data,'' Februar 2019.
  [Online]. Available: \url{http://doi.org/10.5281/zenodo.2553832}
\BIBentrySTDinterwordspacing

\bibitem{Hall/etal/2009a}
M.~Hall, E.~Frank, G.~Holmes, B.~Pfahringer, P.~Reutemann, and I.~H. Witten,
  ``The {WEKA} data mining software: {A}n update,'' \emph{SIGKDD Explorations},
  vol.~11, no.~1, pp. 10--18, 2009.

\bibitem{Chang/Lin/2011a}
C.-C. Chang and C.-J. Lin, ``{LIBSVM}: {A} library for support vector
  machines,'' \emph{ACM Trans. Intell. Syst. Technol.}, vol.~2, no.~3, pp.
  27:1--27:27, May 2011.

\bibitem{Falkenberg/etal/2018a}
R.~Falkenberg, B.~Sliwa, N.~Piatkowski, and C.~Wietfeld, ``Machine learning
  based uplink transmission power prediction for {LTE} and upcoming {5G}
  networks using passive downlink indicators,'' in \emph{2018 IEEE 88th
  Vehicular Technology Conference (VTC-Fall)}, Chicago, USA, Aug 2018.

\bibitem{LeCun/etal/2015a}
Y.~LeCun, Y.~Bengio, and G.~Hinton, ``\BIBforeignlanguage{English (US)}{Deep
  learning},'' \emph{\BIBforeignlanguage{English (US)}{Nature}}, vol. 521, no.
  7553, pp. 436--444, 5 2015.

\bibitem{Quinlan/1992a}
J.~R. Quinlan, ``Learning with continuous classes.''\hskip 1em plus 0.5em minus
  0.4em\relax World Scientific, 1992, pp. 343--348.

\bibitem{Breiman/2001a}
L.~Breiman, ``Random forests,'' \emph{Mach. Learn.}, vol.~45, no.~1, pp. 5--32,
  Oct. 2001.

\bibitem{Cortes/Vapnik/1995a}
C.~Cortes and V.~Vapnik, ``Support-vector networks,'' \emph{Mach. Learn.},
  vol.~20, no.~3, pp. 273--297, Sep. 1995.

\bibitem{Louppe/etal/2013a}
G.~Louppe, L.~Wehenkel, A.~Sutera, and P.~Geurts, ``Understanding variable
  importances in forests of randomized trees,'' in \emph{Proceedings of the
  26th International Conference on Neural Information Processing Systems -
  Volume 1}, ser. NIPS'13.\hskip 1em plus 0.5em minus 0.4em\relax USA: Curran
  Associates Inc., 2013, pp. 431--439.

\end{thebibliography}

\end{document}